\documentclass[12pt]{article}
\usepackage{amsfonts,amssymb,amsbsy,amsmath,amsthm,latexsym,natbib,graphicx}
\usepackage{psfrag,epsfig,epsf} 
\usepackage{subfigure,color,comment}
\usepackage[margin=1in]{geometry}
\usepackage{placeins}

\newtheorem{theorem}{Theorem}
\newtheorem{proposition}{Proposition}

\theoremstyle{definition}
\newtheorem{definition}{Definition}
\newtheorem{assumption}{Assumption}

\newcommand{\blind}{0}

\newcommand{\pihat}{\hat{\pi}}
\newcommand{\Phat}{\hat{P}}
\newcommand{\wbar}{\overline{w}}
\newcommand{\alphabar}{\overline{\alpha}}
\newcommand{\Atil}{\tilde{A}}

\newcommand{\Ptil}{\tilde{P}}
\newcommand{\pitil}{\tilde{\pi}}

\newcommand{\deltatil}{\tilde{\delta}}
\newcommand{\alphatil}{\tilde{\alpha}}

\newcommand{\bmtheta}{\mbox{\boldmath$\theta$}}

\newcommand{\Prob}[1]{\mathbb{P} \left({#1}\right)}

\newcommand{\Abs}[1]{\left\vert{#1}\right\vert}
\newcommand{\Norm}[1]{\left\vert\left\vert{#1}\right\vert\right\vert}
\newcommand{\ind}[1]{1\hspace{-2.3mm}{1}_{\left\{{#1} \right\}}} 
\newcommand{\Expect}[1]{\mathbb{E} \left[{#1}\right]}

\newcommand{\fnq}[1]{q\left({#1}\right)}
\newcommand{\Expects}[2]{\mathbb{E}_{{#1}} \left[{#2}\right]}
\newcommand{\cip}{\stackrel{p}{\longrightarrow}~}
\newcommand{\bmu}{\mathbf{u}}

\newcommand{\bme}{\mathbf{e}}

\newcommand{\md}{\mbox{d}}

\begin{document}

\def\spacingset#1{\renewcommand{\baselinestretch}%
{#1}\small\normalsize} \spacingset{1}


\if0\blind
{
  \title{\bf Adaptive, delayed-acceptance MCMC for targets with expensive likelihoods}
  \author{Chris Sherlock$^1$\footnote{c.sherlock@lancaster.ac.uk}, Andrew Golightly$^2$~and Daniel A. Henderson$^2$}
  \date{{\small $^1$Department of Mathematics and Statistics, Lancaster University, UK\\
      $^{2}$School of Mathematics \& Statistics, Newcastle University, UK \\}}

  \maketitle
} \fi

\if1\blind
{
  \bigskip
  \bigskip
  \bigskip
  \begin{center}
    {\LARGE\bf Title}
\end{center}
  \medskip
} \fi

\bigskip
\begin{abstract}
When conducting Bayesian inference, delayed acceptance (DA)
  Metropolis-Hastings (MH) algorithms and DA pseudo-marginal MH
  algorithms can be applied when it is computationally expensive to
  calculate the true posterior or an unbiased estimate thereof, but a
  computationally cheap approximation is available. A first
  accept-reject stage is applied, with the cheap approximation
  substituted for the true posterior in the MH acceptance ratio. Only
  for those proposals which pass through the first stage is the
  computationally expensive true posterior (or unbiased estimate
  thereof) evaluated, with a second accept-reject stage ensuring that
  detailed balance is satisfied with respect to the intended true
  posterior. In some scenarios there is no obvious computationally
  cheap approximation. A weighted average of previous evaluations of
  the computationally expensive posterior provides a generic
  approximation to the posterior. If only the $k$-nearest neighbours
  have non-zero weights then evaluation of the approximate posterior
  can be made computationally cheap provided that the points at which
  the posterior has been evaluated are stored in a multi-dimensional
  binary tree, known as a KD-tree. The contents of the KD-tree are
  potentially updated after every computationally intensive
  evaluation. The resulting adaptive, delayed-acceptance [pseudo-marginal]
  Metropolis-Hastings algorithm is justified both
  theoretically and empirically. Guidance on tuning
  parameters is provided and the methodology is applied to a
  discretely observed Markov jump process characterising predator-prey
  interactions and an ODE system describing the dynamics of an
  autoregulatory gene network.
\end{abstract}

\noindent%
{\it Keywords:} Delayed-acceptance; surrogate; adaptive MCMC; pseudo-marginal MCMC;
KD-tree.

\spacingset{1.45}

\section{Introduction}
\label{sec:intro}

A major challenge for Bayesian inference in complex statistical models
is that evaluation of the likelihood or, for pseudo-marginal MCMC
\citep{AndrieuR09}, obtaining a realisation from an unbiased estimator
of the likelihood, can be computationally expensive.  The use of a
\textit{surrogate} model with a computationally inexpensive likelihood
in such circumstances has a long history; see, for example,
\cite{SWMW89}, \cite{KennedyO01}, \cite{Rasmussen03},
\cite{BliznyukRSRWM08}, \cite{FieldingNL11}, \cite{Joseph12,Joseph13},
\cite{OverstallW13} and \cite{CMPS:2014}.  The use of inexpensive
surrogates has also been explored in the related context of likelihood
free inference or approximate Bayesian computation (ABC)
\citep{Wilkinson14,MeedsW14}. 

In this paper we propose to use a relatively generic surrogate for
models with expensive likelihoods and we justify its use in adaptive,
delayed-acceptance (pseudo-marginal) MCMC schemes. The delayed
acceptance MCMC algorithm of \cite{ChristenF05} is a two-stage
Metropolis-Hastings algorithm in which, typically, proposed parameter
values are accepted or rejected at the first stage based on a
computationally cheap surrogate for the likelihood. Detailed balance
with respect to the true posterior is ensured by a second
accept-reject step, based on the computationally expensive likelihood,
for those parameter values which are accepted in the first stage.
Delayed acceptance algorithms thus provide draws from the posterior
distribution of interest whilst potentially limiting the number of
evaluations of the expensive likelihood.  Recent examples of the use
of surrogates in delayed acceptance algorithms can be found in
\cite{CuiFO11}, \cite{HigdonRMVF11}, \cite{GolightlyHS15} and
\cite{STG15}, amongst others.  Delayed acceptance algorithms which use
data subsampling and partitioning for tackling large datasets have
also been proposed; see \cite{PayneM14}, \cite{Quiroz15} and
\cite{BGLR:2015}.

For some models there may be an obvious cheap surrogate. For example,
\cite{GolightlyHS15} use both the diffusion approximation and the
linear noise approximation as surrogates for a Markov jump process in
the context of analysing stochastic kinetic models.  For many models,
however, there are no obvious model-based candidates for the
surrogate. It is natural in such scenarios to use regression-based
methods which utilise previous evaluations of the computationally
expensive likelihood to approximate the likelihood or the unnormalised
posterior density at new parameter values. For example,
\cite{BliznyukRSRWM08} use radial basis functions whereas
\cite{Rasmussen03} and \cite{FieldingNL11} use Gaussian processes
(GPs).  
In this paper, we focus on a  
generic surrogate based upon likelihood values at the
$k$-nearest neighbours \citep[e.g.][]{HastieTF09}. It is easy to implement,
computationally cheap, and it adapts as new evaluations of the
computationally-expensive likelihood become available.
In Section \ref{sec.alternatives} we consider the merits and
disadvantages of an alternative, GP-based solution.

We
approximate the likelihood at proposed parameter values by an 
 inverse-distance-weighted average
of the likelihoods of its $k$-nearest neighbours in the training data;
the training data here consist of pairs of parameter vectors and
their corresponding likelihoods.  This $k$-NN approach has
the advantage of being simple, local, flexible to the shape of the
likelihood, and trivially adaptive as new training data become
available.  Our focus on adapting a local approximation is thus
similar to that in \cite{CMPS:2014} and we compare and contrast the
two approaches in Section \ref{sec.alternatives}.

Although it is trivial to update our simple $k$-NN approximation as
new training data become available, a naive implementation of adaptive
MCMC algorithms may not converge to the intended target
\cite[e.g.][]{RobertsRosenthal:2007,AT2008}.  Careful control of the
adaptation is therefore essential and we prove that, subject to
conditions, the strategy we propose is theoretically valid. As with a
number of previous adaptive MCMC algorithms \cite[e.g.][]{RR09,SFR10},
our MCMC kernel is a mixture of a fixed kernel and an adaptive kernel.
Unlike previous such algorithms, however, the adaptive kernel can be
most efficient when the fraction of applications that involve a
computationally expensive evaluation is very low; this impacts on the
rate of adaptation and on the rate of convergence. We therefore also
provide theoretically-justified guidance on choosing both the mixture
probability and the total number of iterations of the algorithm.
 
The main computational expense of the $k$-NN approximation is
searching for the nearest neighbours. In a naive implementation this
search takes $O(n)$ operations, where $n$ is the number of training
datapoints, so  the computational expense of an adaptive, 
nearest-neighbour-based approach would grow linearly with the length
of the MCMC run. Fortunately, there are more efficient
algorithms.  We use an approach based on storing the training data in
a multi-dimensional binary tree, known as a KD-tree
\citep{Bentley:1975,Friedman:1977}. The `K' in `KD-tree'
indicates the dimension of the space, but to avoid confusion with the
number of nearest neighbours $k$, we denote the dimension of the space
by $d$. Because our tree grows on-line as new training data become
available we make two major changes to the standard KD-tree algorithm,
 described at the start of the next section. 
Our adapted KD-tree algorithm allows us to efficiently
search for the nearest neighbours in approximately $O(d\log n)$
operations and add an additional value to the training set also in
$O(d\log n)$ operations.  This yields a highly efficient, adaptive
surrogate.

The remainder of the paper is structured as follows. Our KD-tree
$k$-NN algorithm is described in Section~\ref{sec:kdtree} and its use
in adaptive, delayed-acceptance (pseudo-marginal) MCMC is discussed in
Section~\ref{sec:adaptda}. 
This section also provides the
theoretically-justified guidance on the choice of kernel mixture
probability and number of iterations. Section~\ref{sec:sims} applies
the methodology to a discretely observed Markov jump process
characterising predator-prey interactions and an ODE system describing
the dynamics of an autoregulatory gene network. The paper concludes in
Section~\ref{sec:discuss} with a discussion.

\section{The KD-tree $k$-nearest neighbour algorithm}
\label{sec:kdtree}

Suppose the true parameter is $\theta\in \mathbb{R}^d$. We wish to
find a cheap approximation $\pihat_c(\theta)$ to $\pi(\theta)$ using a weighted
average of the expensive values that have already been
calculated. These expensive values might be of the true posterior
$\pi(\theta_1),\dots,\pi(\theta_m)$ (involving, for example, the numerical
solution of a number of differential equations as in Section~\ref{AR}) or the
expensive values might be
$\pihat_s(\theta_1),\dots,\pihat_s(\theta_m)$, unbiased stochastic
estimates obtained as part of a pseudo-marginal MCMC algorithm
(Section~\ref{LV}).

We will average the $k$-nearest neighbours. 
Naive use of a vector of $n$ $\theta$ values and associated
log-likelihoods, $v$, is expensive. While adding to the list takes
$O(1)$ operations, searching the list for the $k$ nearest neighbours to a
particular point, $\theta^*$, takes $O(n)$ operations.

For our applications, typically the dimension, $d$, of the problem is
moderate: between $3$ and $12$. For very low dimensional problems an
analogue of the quadtree \cite[][]{FinkelBentley:1974} would be the most
efficient approach, but the number of pointers from each node grows
exponentially with dimension. 

We therefore use a variation on the KD-tree
\cite[][]{Bentley:1975,Friedman:1977}. Creation of the tree from $n$
 values takes $O(d n\log n)$ operations and requires storage of
 $O(n)$. For a balanced (see Section~\ref{bal}) tree the computation required to
 add an additional value is $O(d \log n)$, and to search for a nearest
 neighbour is also $O(d\log n)$. 

The standard KD-tree has a single item of data ($\theta$ value and
associated information) at each node; all items further down the tree from
this node have been split at this node as described in detail in
Section \ref{sect.addnode}.  The
standard structure assumes that all of the $\theta$ values to be used are available before the tree is constructed,
whereas our tree potentially grows with each new position at which the
log-likelihood (or an  unbiased estimate thereof) is evaluated. 
The most efficient
 `splitting point' of any node of a tree is the median of all relevant
 values. Use of the median leads to a balanced tree where all end
 nodes would be at approximately the same depth. However we do not
 know the true median. We therefore separate our tree in to
 \textit{leaf nodes} and \textit{branch nodes}. Branch nodes provide
 the splitting information and leaf nodes, which occur at the base of
 the tree, store multiple data
 values. When a leaf node becomes full, it splits according to the
 median of the relevant data values it contains, rather than a true
 median, and becomes a branch
 node, creating two leaf nodes beneath it. The maximum size of a leaf
 node is defined so
 that the leaf median is unlikely to be `too far' from the true
 median; the choice of this tuning parameter 
 is investigated in Section \ref{bal} and in the simulation study in
 Section \ref{LV}.

\subsection{Preliminary run}
\label{sec.prelimrun}
To estimate the likelihood at some new point, we will take a
weighted average of likelihoods, or estimated likelihoods at the $k$
nearest points. If the likelihood varies more quickly with distance
along some axes than along others then `nearest' should be according to
some alternative metric such as a Mahalanobis distance. However we
also divide the KD-tree along hyperplanes which are perpendicular to
one of the Cartesian axes. This leads to gross inefficiencies when
there is a strong correlation between components of $\theta$. For
example if $\theta$ has a bivariate Gaussian distribution with marginal
variances of $1$ and correlation $0.999$ then the hyperplane splits
in a tree of depth $5$, say, effectively partition the first principal component but fail to 
partition the second, so that two $\theta$ values which are reasonably
close according to the Mahalanobis distance can be in different
portions of the tree. This problem is exacerbated in higher dimensions.

The algorithm should, therefore, be more efficient
 if the the parameters are relatively
uncorrelated and if the length scales in the direction of each axis
are similar.
A preliminary run of the MCMC algorithm allows us to find
an approximate center $\hat{\mu}$, and variance matrix
$\hat{\Sigma}$. For all $\theta$ we then define
$\psi:=\sqrt{\hat{\Sigma}}^{-1}(\theta-\hat{\mu})$ to
normalise. 
Since this transformation gives a one-to-one mapping 
from $\theta$ to $\psi$, 
in what follows, for notational simplicity, we refer to the parameter 
values to be stored in the KD-tree as $\theta$ and implicitly 
assume that, in practice, the transformation has already been
applied. 
The preliminary run should be of sufficient length that 
the
sample approximately represents the gross relationships in the main
posterior and hence that 
Euclidean distance is a reasonable metric for the
transformed parameters; the preliminary chain does not need to have
mixed thoroughly. 

\subsection{Tree preliminaries}
\label{sect.treeprel}
The KD-tree stores a large number of vector parameter values,
$\theta\in\Theta$, each with an associated vector of interest, $v_{\theta}\in\mathbb{V}$, in
such a way that the time taken to either update the tree or to
retrieve the information we require is logarithmic in the number of
$(\theta,v_{\theta})$ pairs that are stored in the tree. 
In our case $v_{\theta}=(l_{\theta},n_{\theta})\in\mathbb{R}\times\mathbb{N}$ is the logarithm 
of the average of all estimates of the posterior at parameter
values close to $\theta$ and the number of such estimates.

Associated with the vector of interest is a \textbf{merge} function 
$M:\Theta\times \mathbb{V}\times\Theta\times \mathbb{V}\rightarrow
\mathbb{V}$, which combines the current vector of interest with the vector at a new
value, $\theta^*$.

\subsection{Tree structure}
The tree consists of \textbf{branch nodes} and \textbf{leaf
  nodes}. Each branch node has two children, each of which may either
be a branch node or a leaf node.

\textbf{Leaf nodes}. Each leaf node stores up to $2b-1$
$(\theta,v_{\theta})$ pairs, and the dimension, $d_{split}\in\{1,\dots,d\}$, on which the node will
split. 

\textbf{Branch nodes}. Each branch node stores a component,
$d_{split}\in\{1,\dots,d\}$ on which it was split, and a corresponding scalar value
$\theta_{split}$, the split point. The left-hand child of the node contains (if it is a
leaf node) or points
to (if it is a branch) all $(\theta,v)$ pairs
that have passed through this branch node
 and that have $\theta_{d_{split}}< \theta_{split}$. The
right-hand child contains or points to all $(\theta,v)$ pairs with
$\theta_{d_{split}}>\theta_{split}$. If
$\theta_{d_{split}}=\theta_{split}$ it may be contained by the right
hand node or the left hand node.

\textbf{Root node}. The node at the very top of the tree is called the
root node. By default the root node has $d_{split}=1$. If
there are fewer than $2b$ leaves in the tree then the root node is a
leaf node, otherwise it is a branch node.  

\subsection{Adding data to a tree}
\label{sect.addnode}
When a leaf has $2b$ entries it immediately spawns
 two leaf node children by splitting all $2b$ entries along the
component, $d_{split}$.  The splitting
point, $\theta_{split}$, is the median of the $2b$ values for
$\theta_{dsplit}$. The parent node then
becomes a branch, and the leaf nodes inherit 
$d_{split}=d_{split}^{(p)}\oplus 1$, where $d_{split}^{(p)}$ is
the component on which the parent has just been split, and $\oplus$
represents regular addition except that $d\oplus 1=1$.

Some initial number, $n_0$ of data points are used to create a
balanced tree using the standard recursive procedure given in the 
supplementary material (Appendix~A.1). After this, a
 single new entry $(\theta,l_{\theta})$ is added to the tree by descending from the
root node and, at each branch, comparing component
$\theta_{d_{split}}$ with $\theta_{split}$ to choose the relevant
child (if $\theta_{d_{split}}=\theta_{split}$ then the left child is chosen
 with probability $0.5$). The new entry is added to the leaf node that is reached. If
this leaf node still has fewer than $2b$ entries then the algorithm stops,
otherwise the leaf-node becomes a branch and spawns two child leaf-nodes as
described above.

\subsection{Searching the tree}
We apply a standard two-stage recursive algorithm to find the $r$ nearest neighbours of a given point,
$\theta^*$, together with their distances from $\theta^*$. We
provide an overview of the algorithm below; the procedure is detailed
in full in the supplementary material (Appendix~A.2).  

The algorithm first descends the tree to find the
leaf node to which $\theta^*$ would belong, as described in 
Section \ref{sect.addnode}. It then gradually ascends the tree
from this node to the root node; after each ascent from a node to its
parent a test is conducted to see whether the other child, that is the
child of the current node from which the algorithm has
\textit{not} just ascended, or any of its offspring might hold
 a closer
neighbour than the current $k$ nearest. If this is the case then,
before any further ascent can take place, the search descends down the
tree via this other child.

\subsection{Restricting the growth of the tree}
\label{sect.restric.growth}
The tree will be used to obtain computationally cheap estimates of the
posterior density. We would like to ensure that the accuracy of the
approximations increases with the amount of information stored in the
tree, whether the information arises from exact evaluations of the
posterior or from stochastic estimates. We would also prefer the
cost of obtaining this information to increase slowly,
if at all, with the amount of information. 

To reduce the total number of leaves we set a minimum
distance between leaves, $\epsilon$. For any new information, $(\theta^*,v_{\theta^*})$, to be
added to the tree we first ascertain
whether or not any pairs $(\theta,v_{\theta})$ exist with
$\Norm{\theta^*-\theta}<\epsilon$. If one or more such pair exists
then the merge function, $M$, described in Section \ref{sect.treeprel} is used to combine
$(\theta^*,v_{\theta^*})$ with the nearest pair, providing a
replacement value for $v_{\theta}$; otherwise
$(\theta^*,v_{\theta^*})$ is added to the tree as described in Section \ref{sect.addnode}. 

When each $v_{\theta^*}$ contains an exact evaluation of the posterior
then the merge simply ignores the new information.
However, when $v_{\theta^*}$
contains a stochastic estimate of the posterior some
 weighted average
of the new estimate and of the current average will be more appropriate
since, by the continuity of the posterior, for sufficiently small
$\epsilon$, $\pi(\theta^*)\approx
\pi(\theta)$ for all $\theta^*$ such that
$\Norm{\theta^*-\theta}<\epsilon$. 
In particular therefore, for pseudo-marginal algorithms, we define
\[
M_{PM}(\theta,[l_\theta,n_{\theta}],\theta^*,[l_{\theta^*},1])
:=
\left[\log\left[n_\theta e^{l_{\theta}}+e^{l_{\theta^*}}\right]-\log(n_\theta+1),~n_\theta+1\right].
\]
This vector replaces the previous $[l_{\theta},n_{\theta}]$ vector and
so is associated with the position $\theta$.

\subsubsection*{Choice of merge distance}
Consider a tree with $n$ existing points, 
$\theta_1,\dots \theta_n$ and to which it is proposed that a new point $\theta^*$,
chosen at random, will be added.
We relate the merge
distance, $\epsilon$, to
the probability, $p_{keep}$, that none of the existing points is within the
$\epsilon$ ball of $\theta^*$, so that $\theta^*$ will be added to the
tree. This then provides a guide to setting
$\epsilon$ itself. 

Let $B^*_\epsilon$ be the $\epsilon$ ball around $\theta^*$, define $N_{n,\epsilon}$ to be the number of the
$n$ existing points that are inside $B^*_{\epsilon}$ and consider
$E_{n,\epsilon}=\Expect{N_{n,\epsilon}}$.
The following is proved in the supplementary material (Appendix~B). 
\begin{proposition}
\label{prop.boundpkeep}
If $0<E_{n,\epsilon}<1$ then $1-E_{n,\epsilon}<p_{keep}<e^{-E_{n,\epsilon}}$.
\end{proposition}
To use Proposition \ref{prop.boundpkeep} we require an expression for
$E_{n,\epsilon}$, and this depends on the distribution of $(\theta_1,\dots,\theta_n,\theta^*)$.
For tractability, and because it will often hold approximately with
reasonably sized data sets,  we suppose
 that the target is Gaussian, so that $\theta_i \sim N(\mu,\Sigma)$,
 marginally. This is then normalised (see Section \ref{sec.prelimrun})
 so that the following result (see again Appendix~B for a proof) can be applied.

\begin{proposition}
\label{prop.EnepsGauss}
Let jointly distributed 
$\theta_1\sim N(0,I_d),\dots, \theta_n\sim N(0,I_d)$, 
be independent of $\theta^*\sim N(0,I_d)$. Then 
$E_{n,\epsilon}=nF_{\chi^2_d}(\epsilon^2/2)$, 
where $F_{\chi^2_d}$ is the cumulative distribution function of a
$\chi^2_d$ random variable. 
\end{proposition}
In the simulation studies of Section \ref{sec:sims} we choose $\epsilon$ such that $E_{n,\epsilon}=0.5$, giving
$0.5<p_{keep}<0.61$ and $\epsilon\approx\sqrt{2q_{\chi^2_d}\left({1}/{2n}\right)}$,
where $q_{\chi^2_d}$ is the quantile function of a $\chi^2_d$ random variable.

\subsection{Ensuring that the tree remains balanced}\label{bal}
\label{sect.balance}
We consider two mechanisms through which a KD-tree that is constructed
on-line using an MCMC algorithm may become unbalanced, and for each
problem we provide a solution.

When $d=1$, there are two binary tree structures that allow
for online rebalancing: the red-black tree and the AVL
tree \cite[e.g.][]{Storer:2002}. Unfortunately there are no known algorithms for rebalancing a
KD-tree online and so we consider an alternative which ensures that
the KD-tree remains approximately balanced as at grows.

The splitting hyper-planes define a partition of $\Theta$.
Let the \textit{node box} corresponding to a particular node be the
(possibly unbounded) subset of $\Theta$ defined through the
constraint at each splitting hyper-plane on the journey
 from the root node  to the node in question, as described in Section \ref{sect.addnode}. 

Consider, informally, for any node box, an `effective width' along a particular
co-ordinate axis to be some representative width such as the standard
deviation of the posterior restricted to the node box.
Suppose that the MCMC chain is currently 
in a node box where an effective width  along its splitting
co-ordinate is $\delta$.  
Now, imagine that at each iteration the chain barely moves compared to $\delta$,
and each jump proposal - which can give a new evaluation of the
log-likelihood even if the chain does not move - is also small
compared to $\delta$. In this case all of the new samples for a large number of
iterations will descend to this 
one particular leaf node, which will fill up and then split. 
This split will not, however, be representative of the marginal median
 (in terms of the posterior) for the node box along the splitting axis.
Hence, over the rest
of the MCMC run, once the chain has moved on, there will be one side
of the split which takes most of the future sample points and one side which
takes very few of them, leading to an unbalanced tree.

Now suppose that a preliminary run of $n_0$ iterations has been
carried out and that over this run the chain has been seen to mix
reasonably across the posterior. Let us now construct a
\textit{balanced} tree from this run. Each leaf node will have $b^*$
(or $b^*+1$) entries, for some $b^*\in\{b,b+1,\dots,2b-2\}$. Consider
the node box of any specific leaf node in this tree. Since the
chain has mixed reasonably, it should represent the posterior within this
node box, and in particular (1)
in the component over which the leaf node will split, the median
  should be reasonably approximated, and (2)
in the main MCMC run, the chain should also cover this box in
  approximately $b^*$ iterations or fewer; hence it will also cover any
  sub-divisions of the box in approximately $b^*$ (or fewer) iterations.

The second reason a tree can become unbalanced is Monte Carlo
error. A leaf splits after it has $2b$ entries by finding the median
of $\theta_{d_{split}}$ over the $2b$ entries, but this sample median will
not be the true median over the
node-box. Table \ref{table.treedepthtest} provides, for 4 different values of $2b$,
the probability that the estimated median will be at a true quantile
which is outside the range shown. A tree with $2b\times 10^5$ iid entries was
simulated for each value of $2b$ and each of two dimensions. The table also shows the mean depth of the leaf nodes and
the range of depths of $99\%$ of the leaf
nodes and, in brackets, of all leaf nodes. Both aspects of Table
\ref{table.treedepthtest} suggest that $2b=20$ or $2b=30$
should lead to a reasonably balanced tree with few leaf nodes requiring
much more effort to reach than the majority of the leaf nodes and with
little effect on the overall mean amount of effort required to reach a
leaf node.
\begin{table}[t]
\begin{center}
\begin{tabular}{l|lll|ll}
2b&[0.4,0.6]&[0.3,0.7]&[0.2,0.8]& $d=3$& $d=10$\\
\hline
10&0.49&0.15&0.02    & 18.3, 14-23 (11-25) & 18.4, 14-23 (11-28)\\
20&0.35&0.05&0.002 & 17.7, 15-21 (13-23) & 17.7, 15-21 (12-23)\\
30&0.26&0.02&0.0002 & 17.5, 15-20 (14-22) & 17.5, 15-20 (14-21)\\
40&0.19&0.007&0.00001 & 17.4, 15-19 (14-21) & 17.4, 15-19 (13-20)
\end{tabular}
\caption{Left: the probability that an estimated median from a sample of
  size $2b\in\{10,20,30,40\}$ will be at a quantile outside of the range
  $[0.4,0.6]$, $[0.3,0.7]$ or $[0.2,0.8]$. Right:
    tree depths when $2b\times 10^5$ independent entries are added 
    sequentially to a tree: mean over all leaf nodes, range of the central
    $99\%$ of leaf nodes and the maximum and minimum. \label{table.treedepthtest}}
\end{center}
\end{table}

\section{Adaptive MCMC algorithm}
\label{sec:adaptda}

We briefly review the delayed-acceptance 
algorithm before describing our adaptive version. This is further
extended to an adaptive 
pseudo-marginal version in Section \ref{sect.pseudo.extend}.

\subsection{Delayed-acceptance algorithms}
\label{sect.mcmc.background}
Given a current parameter value, $\theta\in\Theta$, the
Metropolis-Hastings (MH)
algorithm proposes a new value, $\theta^*$ from some density
$q(\theta^*|\theta)$ and then accepts or rejects according to 
\begin{equation}
\label{eqn.stdMHalpha}
\alpha_{MH}(\theta,\theta^*):=1\wedge \frac{\pi(\theta^*)q(\theta|\theta^*)}{\pi(\theta)q(\theta^*|\theta)}.
\end{equation}

The delayed acceptance Metropolis-Hastings (daMH) algorithm utilises a cheap (deterministic or
stochastic) approximation $\pihat_c$ in two stages. At Stage One,
$\pihat_c$ is substituted for $\pi$ in the standard MH acceptance formula:
\begin{equation}
\label{eqn.DAaccOne}
\tilde{\alpha}_1(\theta,\theta^*):= 
1 \wedge
  \frac{\pihat_c(\theta^*)  ~\fnq{\theta | \theta^*}}{\pihat_c(\theta)
  ~\fnq{\theta^*|\theta}},
\end{equation}
A second accept/reject stage is applied to any proposals that pass
Stage One and a proposal is only accepted if it passes both stages. The Stage Two acceptance probability is:
\begin{align}
\label{eqn.DAaccTwo}
\tilde{\alpha}_2(\theta,\theta^*)&:= 
1\wedge \frac{\pi(\theta^*)\pihat_c(\theta)}{\pi(\theta)\pihat_c(\theta^*)}.
\end{align}
The overall acceptance probability,
$\alphatil_1(\theta,\theta^*)\alphatil_2(\theta,\theta^*)$ ensures
that detailed balance is satisfied with respect to $\pi$; however if a
rejection occurs at Stage One then the expensive evaluation of
$\pi(\theta)$ at Stage Two is unnecessary.

\subsection{Adaptive, delayed-acceptance algorithm}
\label{sec.adap.da}
As in \cite{RobertsRosenthal:2007,RR09} our adaptive kernel consists of a mixture of a
fixed kernel and an evolving kernel. At each iteration, 
with a user-defined probability $\beta\in(0,1)$, the fixed
kernel is selected,
otherwise the adaptive kernel is used. Our fixed kernel is a
standard Metropolis-Hasting kernel and our evolving kernel uses
delayed-acceptance  with an acceptance
rate derived from a cheap approximation
$\pihat_c(\theta^*)$ which is an inverse-distance-weighted average of the
expensive evaluations of the true posterior at the $k$ nearest
neighbours to $\theta^*$ in the KD-tree at iteration $n$. 
After iteration $n-1$ let there have been $i_{n-1}$ evaluations of
the true posterior, $\pi$. Let these evaluations be at values
$\theta^*_{i_1},\dots,\theta^*_{i_{n-1}}$.
Our adaptive algorithm requires a sequence of probabilities,
$\{p_i\}_{i\in\mathbb{N}}$, with 
\begin{equation}
\label{eqn.cond.ps}
\lim_{i\rightarrow  \infty}p_i=0.
\end{equation}

In practice, the algorithm proceeds until some $n_{tot}$ iterations
have been performed.

\textbf{Algorithm 1}: \textit{adaptive-KD-tree, delayed-acceptance
  Metropolis-Hastings.}

\begin{enumerate}
\item With probability $\beta$ go to Step \ref{step.MH} (MH)
  else go to
  Step \ref{step.DAMH} (da-MH). 
\item \label{step.MH}
\textbf{MH}: Propose $\theta^*$ from $q(\theta^*|\theta)$. 
Evaluate the expensive posterior,
  $\pi(\theta^*)$, and accept the proposal ($\theta\leftarrow \theta^*$) with
  probability given by (\ref{eqn.stdMHalpha}); otherwise reject the
  proposal ($\theta\leftarrow \theta$). Go to Step \ref{step.adapBMH}.
\item \textbf{da-MH}: Propose $\theta^*$ from $q'(\theta^*|\theta)$.
\label{step.DAMH}
\begin{enumerate}
\item
\label{step.StageOneMH} \textbf{Stage 1}:
Evaluate $\pihat_a(\theta^*)$ using the current KD-tree; with
probability $\tilde{\alpha}_1(\theta,\theta^*)$ as defined in
(\ref{eqn.DAaccOne})  proceed to Step \ref{step.adapSt2MH} (Stage 2); otherwise
reject the proposal ($\theta\leftarrow\theta$), set $i_n=i_{n-1}$, go to next iteration.
\item \textbf{Stage 2}: 
\label{step.adapSt2MH}
Evaluate $\pi(\theta^*)$; accept the proposal ($\theta\leftarrow \theta^*$) 
with probability $\tilde{\alpha}_{2,s}(\theta,\theta^*)$ as defined in
(\ref{eqn.DAaccTwo}); otherwise reject the proposal ($\theta\leftarrow
\theta$). Go to Step \ref{step.adapBMH}.
\end{enumerate}
\item \label{step.adapBMH} Set $i_n=i_{n-1}+1$; add
  $\left(\theta^*,\pi(\theta^*)\right)$ to a list of
  recently-evaluated 
parameter/posterior pairs; with probability $p_{i_n}$ transfer
  all pairs from this list to the KD-tree; go to next iteration.
\end{enumerate}

\subsection{Delayed acceptance random walk Metropolis}
It remains to choose the proposal mechanisms, $q$ and $q'$.
The Random Walk Metropolis (RWM) is a MH algorithm where
$q(\theta^*|\theta)=q(\Norm{\theta^*-\theta})$  for some suitable
norm, and hence $q(\theta^*|\theta)$ and $q(\theta|\theta^*)$ cancel in
the acceptance ratios \eqref{eqn.stdMHalpha} and \eqref{eqn.DAaccOne}. We consider the standard choice of 
\begin{equation}
\label{eqn.std.RWM.prop}
q(\theta^*|\theta)=N(\theta^*;\theta,V)
~~~\mbox{and}~~~
q'(\theta^*|\theta)=q^\xi(\theta^*|\theta):=N(\theta^*;\theta,\xi^2V),
\end{equation}
where $N(\cdot ;\theta,V)$ denotes a multivariate Gaussian density 
with mean $\theta$ and variance $V$ and where  
 $V$ has been chosen so as to approximately optimise the
efficiency of the standard RWM algorithm.

As we shall discover, the cheap approximation, $\pihat_c$ is
reasonably accurate.  
 As $\xi$ increases from $1$ the overall acceptance rate and, in particular,
the Stage One acceptance rate, $\alpha_1$, can decrease quite
substantially. If all computational expense is negligible except for
the evaluation of the true posterior then for a given amount of
computational effort, the number of evaluations of the expensive
posterior remains approximately constant, although the total number of
iterations of the algorithm increases in proportion to the reciprocal of
$\alpha_1$. However, as each proposed jump is larger, moves which
are accepted at Stage Two are typically larger. Thus, provided the
Stage Two acceptance rate does not decrease too drastically, the
mixing of the algorithm (in terms of movement per CPU second) can, and
often does, actually increase for intermediate values of $\xi$.
This
heuristic has been noted before \cite[e.g.][]{ChristenF05,BGLR:2015}
and also applies for the delayed-acceptance pseudo-marginal
RWM. A rigorous analysis of the behaviour of these algorithms as a function of the
scaling is provided in \cite{STG15}.

\subsection{Adaptive, delayed-acceptance, pseudo-marginal algorithm}
\label{sect.pseudo.extend}
We first overview pseudo-marginal Metropolis-Hastings algorithms and
then describe the adjustments to the set-up required for the
pseudo-marginal version of our algorithm. The algorithm itself is
provided in the supplementary material (Appendix~C).
 
The pseudo-marginal algorithm uses a non-negative stochastic estimator 
$\hat{\pi}_s(\theta;Z)$ of the posterior $\pi(\theta)$, where 
 $Z$ is a collection of random variables whose
distribution may, and usually does, depend on $\theta$. Crucially, we
require $\Expect{\pihat_s(\theta;Z)}=c\pi(\theta)$, where $c$ is fixed and
non-negative. We may therefore rewrite $\pihat_s(\theta)$ as $\pi(\theta)W$ 
 with $W\in \mathcal{W}\subseteq [0,\infty)$ sampled from some density
 $q_\theta(w)$, and 
\begin{equation}
\label{eqn.expect.W}
\Expect{W}=\int_{0}^\infty wq_{\theta}(w)~\md w=c.
\end{equation}
 The pseudo-marginal MH (PsMMH) algorithm is
 simply a Metropolis-Hastings Markov chain acting  on the
 extended statespace $\Theta \times \mathcal{W}$ with a target of
\begin{equation}
\label{eqn.extended.stat.dens}
\pitil(\theta,w)=\frac{1}{c} \pi(\theta)q_{\theta}(w)w.
\end{equation}
This has the required marginal for $\theta$ by \eqref{eqn.expect.W}.
Detailed balance is ensured with respect to this target by setting the
probability for $(\theta^*,w^*)$ being accepted to:
\begin{equation}
\label{eqn.stdPMMHalpha}
\alpha_{PM}([\theta,W],[\theta^*,W^*]):=1\wedge
\frac{\pihat_s(\theta^*)q(\theta|\theta^*)}{\pihat_s(\theta)q(\theta^*|\theta)}
=1\wedge \frac{\pi(\theta^*)q(\theta|\theta^*)W^*}{\pi(\theta)q(\theta^*|\theta)W}
.
\end{equation}

When delayed acceptance is implemented with a pseudo-marginal
framework, the Stage One acceptance probability is exactly as in
\eqref{eqn.DAaccOne}. In Stage Two the true posterior in
\eqref{eqn.DAaccTwo} is replaced with the realisation from the
unbiased estimator:
\begin{align}
\label{eqn.PMDAaccTwo}
\tilde{\alpha}_{2,PM}(\theta,\theta^*)&:= 
1\wedge \frac{\pihat_s(\theta^*)\pihat_c(\theta)}{\pihat_s(\theta)\pihat_c(\theta^*)}.
\end{align}

The kernel $\Ptil$ is now a fixed PsMMH kernel on
$\Theta\times \mathcal{W}$ and 
$\{\Ptil_\gamma\}_{\gamma\in\mathcal{G}}$ is now  a set of
pseudo-marginal kernels on $\Theta \times \mathcal{W}$. 
The common stationary density of all kernels is now given in
\eqref{eqn.extended.stat.dens}, so, very importantly, all
kernels use the same mechanism for generating the estimate
$\pihat_s(\theta^*)$ of the posterior at the proposed value for
$\theta$. 

The algorithm proceeds as for the non-pseudo-marginal version except that
$\pihat_s$ is substituted for $\pi$ in \eqref{eqn.stdMHalpha} (fixed
kernel), and \eqref{eqn.DAaccTwo} is replaced with
\eqref{eqn.PMDAaccTwo} (evolving, DA kernel). Naturally, instead of storing evaluations of $\pi$ the KD-tree now stores
realisations of the unbiased approximation, $\pihat_s$. 

\subsection{Theory and guidance}
\label{sect.theoryguide}
We show that, subject to conditions, our algorithms (with general proposals, $q$) are ergodic. We 
also provide guidance on choosing the probability of using the fixed
kernel, $\beta$, and on the number of iterations for which the
algorithm should be run.

Define $\alpha_{MH}(\theta,\theta^*)$ as follows. For the adaptive
KD-tree daMH algorithm of Section \ref{sec.adap.da} $\alpha_{MH}(\theta,\theta^*)$ is the
acceptance probability for the fixed kernel as given in \eqref{eqn.stdMHalpha}. For the pseudo-marginal
version of the algorithm in Section \ref{sect.pseudo.extend} it is the acceptance probability for an
hypothetical, idealised version of the fixed, pseudo-marginal kernel where the
posterior is known exactly, up to a fixed multiplicative constant: 
$\alpha_{MH}(\theta,\theta^*)=\alpha_{PM}([\theta,1],[\theta^*,1])$,
where $\alpha_{PM}$ is defined in \eqref{eqn.stdPMMHalpha}. We require
 a minorisation condition and, for the daPsMMH algorithm, and additional
assumption of uniformly bounded weights. These assumptions are
discussed in in the supplementary material (Appendix~D), where their main
consequence, Theorem \ref{thrm.adap.ergod} is proved.

\begin{assumption}
\label{assump.minorisation}
There is a density $\nu(\theta)$ and $\delta > 0$ such that
$q(\theta^*|\theta)\alpha_{MH}(\theta,\theta^*)\ge \delta \nu(\theta^*)$
for all $\theta \in \Theta$.
\end{assumption} 

\begin{assumption} 
\label{assump.wbar}
The support for $W$ is uniformly (in $\theta$) bounded above by
  some $\wbar<\infty$.
\end{assumption} 

\begin{theorem}
\label{thrm.adap.ergod}
Subject to Assumption 1 the adaptive KD-tree daMH algorithm of
Section \ref{sec.adap.da} is ergodic. The adaptive KD-tree daPsMMH
algorithm of Section \ref{sect.pseudo.extend} is ergodic subject to
Assumptions \ref{assump.minorisation} and \ref{assump.wbar}.
\end{theorem}

Now consider the specific, scaled, proposal $q^\xi$ defined in
\eqref{eqn.std.RWM.prop}, where  
increasing $\xi$ decreases the Stage One acceptance rate,
$\alpha_1$. The algorithm may only accept a proposal after a
computationally-expensive 
evaluation (of $\pi$ for daMH, or $\pihat_s$ for daPsMMH).
Thus, if the probability, $\beta$, that the non-DA kernel will be
chosen is unaltered, then as $\alpha_1\rightarrow 0$ nearly all of
the expensive evaluations  will be by the fixed, non-DA kernel, and the
relative contribution from the DA kernels will unintentionally dwindle to zero.

Decreasing $\beta$ in proportion to $\alpha_1$ would fix the
fraction of all expensive evaluations that are by the DA kernel;
however with a smaller $\beta$ the chain can no longer be guaranteed to
 be as close to $\pi$ after the same, fixed number of iterations.
Theorem~2, which is stated and proved in the 
supplementary material (Appendix~D), shows that, with
$\beta\propto \alpha_1$, as $\alpha_1$ decreases the total number
of iterations, $n$, of the algorithm should be increased so as to 
maintain the expected total number of expensive
evaluations of the posterior, $\Expect{I^\xi_{n}}$, and that $\Expect{I^\xi_{n}}$ can be set so as to
maintain any given upper bound on the total variation distance between
the chain and $\pi$ whatever the value of $\alpha_1$. 

Fixing $\Expect{I^\xi_{n}}$ approximately fixes the expected
overall CPU cost. Furthermore, adaptation can only occur when the expensive posterior is
evaluated, and it occurs with a fixed set of probabilities that depend
on $i$ and not on the iteration number. So, fixing 
$\Expect{I^\xi_{n}}$ also approximately fixes the expected number of adaptation occurrences.

\section{Simulation Studies}
\label{sec:sims}

In this section we evaluate the empirical performance of the proposed 
da-PsMMH and da-MH algorithms by considering two examples based upon Markov jump processes
(MJPs). 
The first example (Section \ref{LV}) arises from the Lotka-Volterra system of
predator-prey interactions 
\cite[e.g.][]{boys08}. 
 Since the marginal 
likelihood is intractable, we apply the adaptive da-PsMMH scheme and
compare its 
performance over a range of tuning parameter choices with that of an 
optimised PsMMH scheme. 
The second example (Section~\ref{AR}) arises from the 
autoregulatory network proposed by \cite{Golightly05}. Given the size 
and complexity of this system, a linear noise approximation (LNA)
\citep{kampen2001} (see also Appendix~F of the supplementary material), 
of the corresponding Markov jump process is taken to be the inferential 
model of interest. Following the algorithm of 
\cite{fearnhead12} (see Appendix~F.1), the marginal likelihood under this model is tractable, 
but involves the solution of a system of 14 coupled ordinary differential 
equations (ODEs), which can be time consuming. We therefore apply the 
da-MH algorithm and compare its performance to a simple MH scheme 
without delayed acceptance.

Both MJPs
 are described through a set of
$r$ reactions between $p$ different species,
$\mathcal{X}_1,\dots,\mathcal{X}_p$. The hazard rate of each reaction depends
on the current species numbers, $X_1,\dots,X_p$ via an assumption of
mass-action kinetics with unknown reaction rate constants 
 $\nu_1,\dots,\nu_r$; for further details regarding the construction
of MJP representations of reaction networks we refer the reader to
\cite{Wilkinson12}. Tables~E.1 and E.2 in 
Appendix~E.1 list the reactions and associated hazards for each example. 

For the Lotka-Volterra model, the Gillespie algorithm
(\cite{Gillespie77}) was applied, using parameter values taken from
\cite{Wilkinson12}, to generate a skeleton path comprising 51 values of $X_t$ at integer times 
in the interval $[0,50]$. For the
autoregulatory system,  
the LNA itself was used, with parameter values taken from \cite{golightly11}, to generate two
skeleton paths containing, respectively, $101$ and $201$ values of 
 $X_t$ at evenly-spaced times covering the intervals $[0,100]$ and 
$[0,1000]$. All skeletons were then corrupted with Gaussian noise to
form the data sets on which inference was performed:
\begin{equation}
\label{eqn.Gaussian.noise}
Y_t|X_t=x_t~\sim N(x_t,D),
\end{equation}
where $D$ is a diagonal matrix
with diagonal entries $\sigma_1^2,\dots,\sigma_p^2$. In the
autoregulatory example we refer to the data sets with $101$ and $201$
observations as $\mathcal{D}_1$ and $\mathcal{D}_2$ respectively. 
Appendix~E.1 provides details of the
initial conditions and parameter values for each simulation as well as
for the variances of the corrupting Gaussian noises.  

For inference, in both cases,
for simplicity, the initial state of the system is fixed at its
(known) true value. As all of the
parameters are strictly positive we therefore consider a logarithmic
transformation so that the parameter vector of interest is 
$\theta=\left(\log(\nu_{1}),\dots,\log(\nu_{r})\right.$, $\left.\log(\sigma_{1}),\dots,\log(\sigma_{p})\right)$.
In both examples, individual components of $\theta$ for which
inference is performed are given
independent Uniform $U(-8,8)$ priors. 
For the Lotka-Volterra model, $r=2$ and $p=2$ so that
$\mbox{dim}(\theta)=4$. 
For the autoregulatory system, $p=4$ and, as discussed in Appendix~E.1,  we
 perform inference on $r=6$ rate constants, giving  $\mbox{dim}(\theta)=10$.

Random-walk proposals of the form \eqref{eqn.std.RWM.prop} are used
for the fixed and the adaptive kernels in both examples. The fixed
kernels used a proposal variance of $V_{fixed}=\lambda \hat{\Sigma}$, where
$\hat{\Sigma}$ is the sample variance
from an initial, pilot run using a fixed, non-delayed-acceptance
kernel, and $\lambda$ is chosen to optimise the efficiency of the
fixed kernel. The corresponding adaptive kernels use a proposal variance of
$\xi^2V_{fixed}$, with $\xi>0$ a tuning parameter.

In each example we take the probability of adding to the tree at
iteration $n$ to be
\begin{equation}
\label{eqn.define.dim.probs}
p_{i_n}=(1+ci_n)^{-1}.
\end{equation}
For each example, the
computational cost of an evaluation of the expensive stochastic
approximation was estimated from the initial training run, and this
provided an estimate of the number of expensive evaluations,
$\hat{i}$, that would fit within the computational budget.
Using the guidelines in Section 2.6, the parameter $\epsilon$ was
chosen with a desire that points still be
as likely as not to be added to the tree after $\hat{i}/2$ evaluations had
already been added. For both systems this had the effect
of limiting the overall tree size to around four times the size of the
initial training set.

\subsection{Discretely observed Markov jump process}\label{LV}


To implement the adaptive da-PsMMH 
scheme described in Algorithm~1, we used a bootstrap particle filter with $m$ particles \citep{andrieu10} to obtain each value of 
$\hat{\pi}_{s}(\cdot)$. 
Both $m$ and $V_{fixed}$ were chosen
so as to optimise the efficiency of the fixed kernel; see the supplementary material (Appendix~E.2).
 We initialised 
the KD-tree using the first $10^{4}$ evaluations of the expensive
posterior in the pilot run. For all experiments, we assumed a fixed computational budget of $10^{4}$ 
seconds, which equated to approximately $\hat{i}=40000$ evaluations of
the expensive posterior and, thus, a maximum tree size of five times
the initial size. 

\begin{figure}[ht]
\begin{center}
\psfrag{xi}[][]{$\xi$}
\psfrag{alp1}[][]{$\hat{\alpha}_{1}$}
\psfrag{RelESS}[][]{\small Relative mESS}
\includegraphics[angle=270,width=\textwidth]{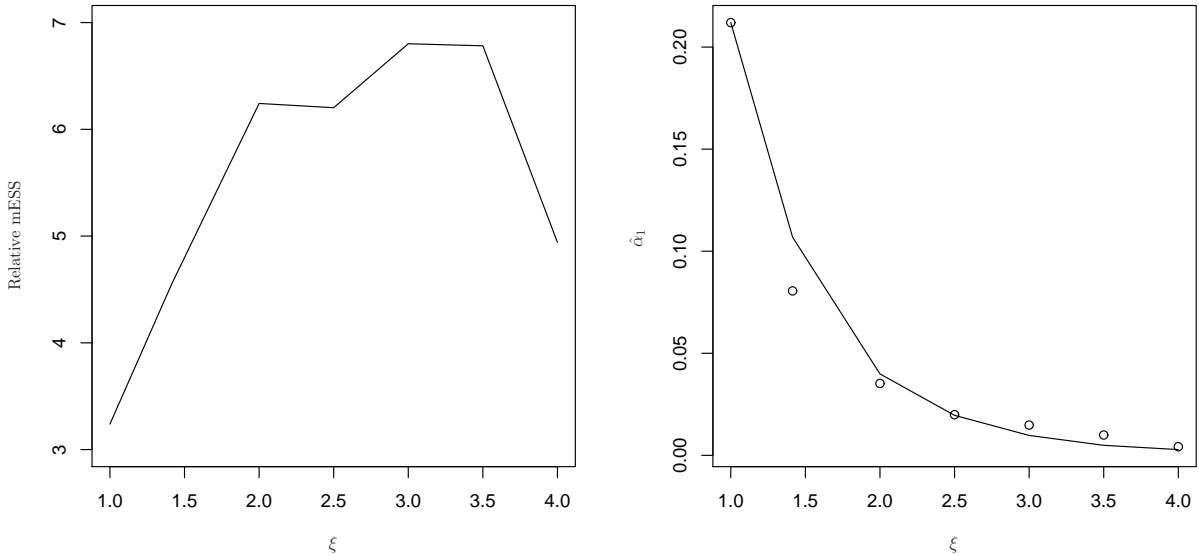}
\caption{Left panel. Minimum effective sample size (mESS) relative to optimised PsMMH, against scaling. 
Right panel. Empirical stage 1 acceptance probability $\hat{\alpha}_{1}$ 
against scaling. The points represent
$\hat{\alpha}_1(\xi=1)\beta(\xi)/\beta(\xi=1)$ and show that $\beta(\xi)$
was scaled in proportion to $\alpha_1(\xi)$. \label{fig.LVess}
}
\end{center}
\end{figure}

To assess the effect of scaling 
$\xi$ on the overall and relative (to PsMMH) efficiency of da-PSMMH we fixed the number 
of leaf nodes in the KD-tree upon splitting to be $2b=20$, the number of nearest 
neighbours to be $k=b=10$ and took the parameter controlling the rate of adaptation to be $c=0.001$ 
to give around a 5\% chance of adaptation after half the computational budget. 
We then ran the algorithm for values of $\xi\in[1,4]$, following the practical advice 
of Section \ref{sect.theoryguide} by choosing $\beta\propto
\hat{\alpha}_{1}$ with
$\beta=0.05$ for $\xi=1$. 
Firstly, the sampled posterior values 
are consistent with the ground truth parameter values that produced
the data (see 
Figure~G.1 in the supplementary material for 
the marginal posterior distributions for a typical run).
 Figure~\ref{fig.LVess} 
shows the effect of scaling on minimum effective sample size (mESS) over each parameter chain 
relative to that obtained under an optimally tuned PsMMH scheme (with an acceptance rate of $9.9\%$ and 
an mESS of 528) and the effect of scaling on the 
Stage 1 acceptance probability. The (scaled) values of $\beta$ used for each run are also shown. 
Figure~\ref{fig.LVess} suggests that for the values of $\xi$ considered, $\xi=3$ is optimal 
in terms of mESS and gives an improvement on overall efficiency over PsMMH of a factor of 6.8. Even simply  
 taking the same scaling as PsMMH still gives a 3-fold increase in efficiency 
of da-PsMMH over PsMMH.   

To assess the effect of adaptation on the performance of the algorithm we fixed $\xi=3$, 
$k=b=10$ and performed runs with $c\in:=\{0.0001,0.001,0.01,\infty\}$ with 
$c=\infty$ representing no adaptation. Table~\ref{table.LVc}
summarises our findings. At $\hat{i}/2$ iterations, these values of
$c$ correspond to an expected number of expensive
evaluations before adaptation occurs of approximately
$\{3,20,200,\infty\}$, respectively. The larger the pause between
adaptations, the less accurate the tree is between adaptations, and
while $200$ new evaluations is small compared with $10000$ or more
existing evaluations, it must be remembered that the most recent
evaluations will be from a similar part of the state space to the
current position and so will be among the most relevant.
The reduction in accuracy especially in new, low-density regions, increases the Stage 1 acceptance rate and 
decreases the Stage 2 acceptance rate 
giving an overall reduction in statistical efficiency. Moreover, the
increase in the Stage 1 rate 
results in a larger number of expensive posterior evaluations. 

With $c=\infty$ the algorithm runs with no adaptation and just uses
the initial training set of $10,000$ posterior evaluations. 
In this case (last row of Table~\ref{table.LVc}) performance is better
than for the simple RWM algorithm but worse than for all of the cases
that allow adaptation, providing clear evidence of the importance of
adaptation for our algorithm. Further, the more slowly $p_i\downarrow \infty$,
the more efficient the algorithm.

We also explore the sensitivity of our method to the choice of the number of leaf nodes 
in the KD-tree upon splitting ($2b$) and the number of nearest neighbours ($k$). We 
fixed $\xi=3$, $c=0.001$ and took 
$2b\in\{4,10,20,30\}$ and $k\in\{2,5,10,15\}$.  Table~G.3 in Appendix~G 
shows empirical performance for each $(k,b)$ combination
considered. Consistent with our findings in Section \ref{sect.balance}, increasing
$b$ increases the efficiency until $2b=20$, but there is little
difference when moving from $2b=20$ to $2b=30$. Fixing $2b$ and varying $k$ suggests that $k=5$ is optimal in terms 
of mESS. Further discussion can be found in the supplementary material (Appendix~G). 
Finally we examine the gain in overall efficiency by using a KD-tree as a storage and look-up 
method over 
simply storing posterior evaluations in a list. Running 
the da-PsMMH scheme with the optimal values ($\xi=3$,
$c=0.001$ and $k=5$) gave an mESS of $2009$. Thus, in this example, 
using a KD-tree increases overall efficiency over a naive approach by a factor of 1.9. 
Naturally, increasing the computational budget (and therefore the number of posterior 
evaluations to be stored) will increase the advantage of the KD-tree.

\begin{table}[ht]
\begin{center}
\begin{tabular}{|c|ccccccc|}
\hline
$c$ & Tree Size & Mean depth & depth range &$\hat{\alpha}_{1}$ & $\hat{\alpha}_{2}$ & mESS & Rel. mESS   \\
\hline
0.0001   &41078 &11.82 &10-14 &0.00772 &0.339 &3845 &7.28 \\
0.001    &40256 &11.79 &10-14 &0.00915 &0.276 &3591 &6.80 \\
0.01     &43248 &12.04 &10-15 &0.0121  &0.204 &2464 &4.67 \\
$\infty$ &10000 &9.69  &9-10  &0.0175  &0.136 &1829 &3.46 \\
\hline
\end{tabular}
\caption{Effect of rate of adaptation $c$. Final tree size, mean leaf node depth, depth range, empirical stage 1 and 2 
acceptance rate, minimum effective sample size (mESS) and relative mESS.
\label{table.LVc}}
\end{center}
\end{table}

\subsection{Discretely observed ODE system}\label{AR}


 The LNA gives a Gaussian model 
for $X_t$ which when coupled with the Gaussian observation model above, permits 
a tractable form for the marginal likelihood which we denote by
$\pi(y_{1:n}|\theta)$. An algorithm for evaluating the marginal likelihood, and 
therefore the  posterior (up to proportionality) under the LNA, can be found in the supplementary material 
(Appendix F.1). Executing one 
iteration of the algorithm requires calculation of a full numerical solution of the ODE system 
(F.1) over $[0,100]$ or $[0,1000]$. Our implementation uses standard routines from 
the GNU scientific library, specifically the explicit embedded 
Runge-Kutta-Fehlberg $(4, 5)$ method. We limit the computational cost of these calculations 
by applying the da-MH scheme. 

The pilot run for each dataset was of $3\times 10^{4}$ iterations.
 We initialised 
the KD-tree with all $3\times 10^{4}$ evaluations of the expensive posterior obtained from the 
initial pilot run. For all experiments, we assumed a fixed computational budget of $5\times 10^{3}$ 
seconds, which equated to approximately $\hat{i}=120000$ evaluations of
the expensive posterior and, thus, a maximum tree size of five times
the initial size. 
 Following the findings of Section~\ref{LV} we initially set the adaptation rate
 to $c=0.001$, and set 
$2b=20$ and $k=5$. 

\begin{figure}[ht]
\begin{center}
\psfrag{xi}[][]{$\xi$}
\psfrag{ESS}[][]{\small mESS}
\psfrag{RelESS}[][]{\small Relative mESS}
\includegraphics[angle=270,width=\textwidth]{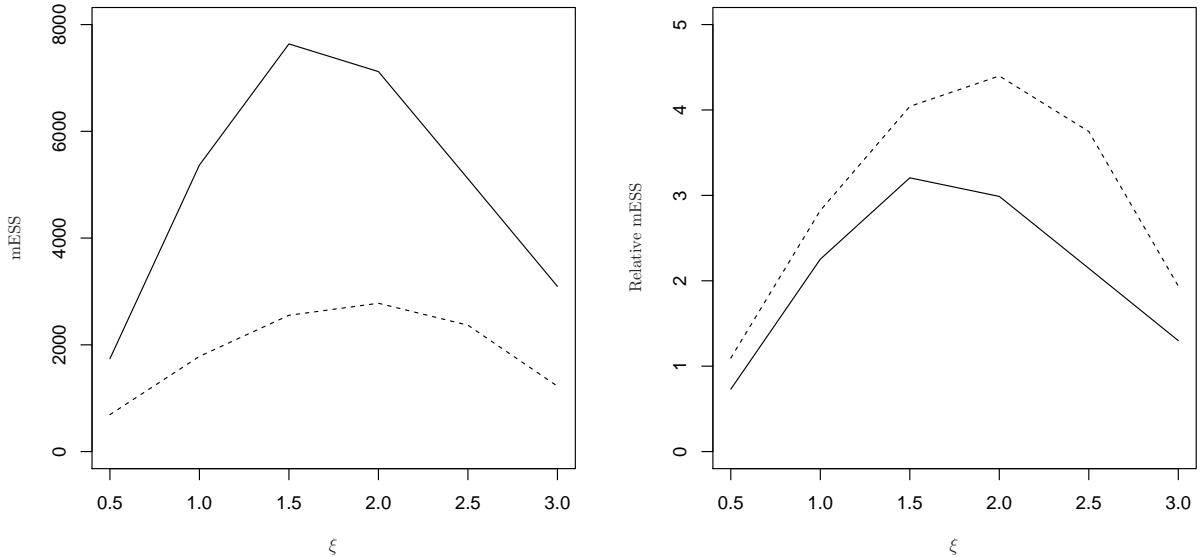}
\caption{Left panel. Minimum effective sample size (mESS) from da-MH against scaling. 
Right panel. Minimum effective sample size from da-MH relative to the same quantity from optimised MH, against scaling. 
For each panel, the solid and dashed lines indicate output using datasets $\mathcal{D}_{1}$ and $\mathcal{D}_{2}$ respectively.\label{fig.ARess}
}
\end{center}
\end{figure}

Firstly,
Figure~G.3 in Appendix~G shows that the sampled 
parameter values are consistent with the ground truth, with a decrease in uncertainty when using 
more observations. With dataset $\mathcal{D}_{2}$ the LNA equations must be solved 
over a longer time period than for $\mathcal{D}_{1}$, and the steps of the algorithm 
for calculating the marginal likelihood (in Appendix~F.1) must be executed twice as 
many times. Consequently Figure~\ref{fig.ARess} (left panel) shows that the minimum effective 
sample size (mESS) obtained under da-MH is smaller when using dataset $\mathcal{D}_{2}$. However, 
mESS relative to the same quantity under MH is increased when using $\mathcal{D}_{2}$, since the 
cost of evaluating the KD-tree is unchanged (for a fixed tree size). The optimal scaling $\xi$ 
(for the values considered) for each scheme is reported in Table~\ref{table.AR}. We 
also report output of additional runs with $c\in:=\{0.0001,0.001,\infty\}$. An 
optimally tuned da-MH scheme (with $c=0.001$) gives an increase in overall efficiency of a factor 
of 3.2 when using $\mathcal{D}_{1}$ and 4.4 when using $\mathcal{D}_{2}$. When 
$c=\infty$ (representing no adaptation), we see an increase in Stage~1 acceptance rate and a decrease 
at Stage~2. The resulting decrease in empirical performance provides further evidence of the 
importance of adaptation. Finally, we again note that the algorithm performs best with a very low value
of $c$ whilst still providing posterior output consistent with $c=0.001$ (results not shown).

\begin{table}
\begin{center}
\begin{tabular}{|c|cccccc|}
\hline
Algorithm & $\xi$ & $c$ & $\hat{\alpha}_{1}$ & $\hat{\alpha}_{2}$ & mESS & Rel. mESS   \\
\hline
& \multicolumn{6}{c|}{$\mathcal{D}_{1}$ (101 obs. on $[0,100]$)}\\
MMH   &1.0&-- &0.225 &1.000 &2383  & 1.00 \\
da-MMH&1.5&0.0001  &0.141 &0.482 &7699  & 3.23\\
      &1.5&0.001   &0.131 &0.480 &7638  & 3.21\\
      &1.5&$\infty$&0.171 &0.366 &4530  & 1.90\\
& \multicolumn{6}{c|}{$\mathcal{D}_{2}$ (201 obs. on $[0,1000]$)}\\
MMH   &1.0&-- &0.2291 &1.000 &632  &1.00 \\
da-MMH&2.0&0.0001  &0.0455 &0.394 &2996 &4.74 \\
      &2.0&0.001   &0.0390 &0.338 &2779 &4.40 \\
      &2.0&$\infty$&0.0978 &0.165 &1485 &2.35 \\
\hline
\end{tabular}
\caption{Algorithm, optimal scaling ($\xi$), adaptation rate parameter ($c$), empirical stage 1 ($\hat{\alpha}_{1}$) and 2 
($\hat{\alpha}_{2}$) acceptance rate, minimum effective sample size (mESS) and relative mESS.
\label{table.AR}}
\end{center}
\end{table}


\section{Discussion}
\label{sec:discuss}

We have presented standard and pseudo-marginal versions of an adaptive, delayed-acceptance 
random walk Metropolis algorithm. The delayed-acceptance (DA) step is
generic, estimating the posterior using an inverse-distance-weighted average of the
$k$-nearest previous evaluations of the posterior, and the search for
these neighbours is made fast by storing a subset of previous
evaluations in a customised version of a KD-tree. The kernel is a
mixture of a fixed (non-adaptive, non-DA) kernel and an adaptive DA
kernel. Easy-to-use \texttt{C} code for creating a KD-tree and storing
and retrieving values from it is provided alongside this article.

We have shown that our algorithm is ergodic, subject to
conditions. Furthermore, as the scaling of the RWM proposal in the
DA kernel is increased, the probability of choosing
the fixed kernel should be decreased in proportion to the Stage One
acceptance rate of the DA kernel and the total number of
iterations should be scaled so that the expected number of 
evaluations of the expensive posterior remains constant or, 
equivalently, so that the total computational budget remains fixed.

Pseudo-marginal and non-pseudo-marginal versions of the methodology
were applied, respectively, to synthetic data generated from a
discretely observed Lotka-Volterra system and an ODE model of species
dynamics in an autoregulatory gene network.  
In these two examples, our proposed scheme outperforms the standard scheme
by factors of approximately 7 and 4 for, respectively.

In both of our examples the algorithm was more efficient the more
slowly the adaptation probability approached zero, suggesting that it
might be best to simply add each new evaluation of the expensive
posterior directly to the KD-tree directly rather than storing them in
a queue. Forcing the limit of the adaptation probability to be $0$ ensures the
diminishing adaptation condition (see, e.g., Theorem~5) is satisfied; diminishing adaptation is itself one
of the key conditions required for ergodicity and the concern would be
that by removing this direct constraint the algorithm would no longer
be ergodic. Whilst this seems likely to be the case in general, 
Theorem~3 in the supplementary material (Appendix~D.5.1) shows that for a 
variation of Algorithm 1 on a compact state space it is possible to
adapt after (at worst) every other expensive iteration and still remain ergodic. 

We have focussed on making the cheap approximation to the posterior
adaptive by intermittently updating the KD-tree. The covariance matrix
of our random walk proposal could have been made adaptive by
intermittently updating the covariance matrix of the random walk
proposal so that it uses all entries in the chain to date
\cite[e.g.][]{RR09,SFR10}. This could even be made `local', using only the $k$-nearest
neighbours in the KD-tree. It might also be possible to adaptively
update the scaling of the
proposal; however the mechanism to use is less obvious since, unlike
in \cite[e.g.][]{AT2008,SFR10,Vihola2012}, there is
no single optimal acceptance rate for our algorithm. Such
adaptations would be a distraction to our main innovation 
and  have not been implemented.

As we were revising this article, \cite{KostovWhitely2016} proposed
an algorithm for estimating the variance of the estimator of the
likelihood that comes from the particle filter. Although we do not
pursue it here, this opens up the possibility of weighting each
estimate of the likelihood according to the reciprocal of its variance
as well as its distance from the proposed $\theta^*$.

\subsection{Alternatives to k-nearest neighbours}
\label{sec.alternatives}

As mentioned in Section~\ref{sec:intro}, there are similarities
between the non-pseudo-marginal version of our algorithm and that of
\cite{CMPS:2014} (henceforth denoted CMPS). Here we discuss the
approach of CMPS, highlighting both similarities to and differences
from our algorithm.  We also consider a general GP alternative
to our k-NN implementation. 

CMPS fit local linear-, quadratic- and
Gaussian-process (GP)-based models in neighbourhoods of candidate
values, and at any given point in time their algorithm targets the
approximate posterior rather than the true posterior. However, the
accuracy of the approximation is continually assessed and improved by
carefully choosing further local points as the algorithm proceeds so
that, asymptotically, the algorithm targets the true posterior
distribution. Between adaptations our algorithm targets the true
posterior distribution, but it is perturbed every time an adaptation
occurs. Thus our algorithm also asymptotically targets the true
posterior distribution. Both algorithms also use an increasing set of
evaluations of the true posterior.  CMPS chooses the next evaluation
of the true posterior by design whereas our algorithm is
``opportunistic'' and potentially adds the value at each new point
evaluated; however it is an intelligent opportunist in the detail of
how it deals with new points which are very close to existing points.

Our algorithm could be extended to fitting local planes, quadratics or
GPs to the $k$-nearest neighbours in a similar manner to CMPS, however
the weighted average is simpler, quicker, and for it to be a sensible
approach to take it only requires the posterior to be bounded and
continuous, rather than needing additional constraints.  When it is
the exact log-likelihood that we are approximating these more complex
local models might lead to an improvement in the accuracy of the
surrogate, at the small expense of possibly having to use more nearest
neighbours to fit.  In the pseudo-marginal case, however, the
efficiency will depend on the variance in the log-likelihood estimates
and the true changes in log-likelihood. If the former outweighs the
latter then there is little to be gained as the extra computation may
not lead to additional accuracy. The cost of re-estimating the GP
hyper-parameters as new training data become available is discussed in
the final paragraph of this section. One final difference between the two approaches is that
CMPS focusses on the non-pseudo-marginal case whereas our algorithm is
applied in both pseudo-marginal and non-pseudo marginal settings.

Our likelihood estimate at a proposed point, $\theta'$, is a weighted average of the likelihoods at the k-nearest neighbours. An alternative approach would be to use a Gaussian process (GP) fitted to the set of $(\theta,\log p(y|\theta))$ pairs currently in the kd-tree. GPs have been used to approximate the log-likelihood previously. For example in \cite{Rasmussen03} and \cite{FieldingNL11} a GP provides a cheap surrogate for Hamiltonian Monte Carlo calculations. Alterntively, in the context of ABC-MCMC, \cite{Wilkinson14} uses a Gaussian process to model the logarithm of the ABC approximate likelihood function, while \cite{MeedsW14} approximate the joint synthetic likelihood at the current and proposed point using independent GPs for each summary statistic.

As with our approach, the final point estimate from a GP is a weighted average of existing values. However,  
 by estimating the parameters of the GP one may represent the 
scales of variability more accurately and so obtain more accurate point estimates than with our inverse-distance weighting approach. The GP model also supplies an estimate of the uncertainty in the point estimate of 
the log-likelihood.

As in \cite{CMPS:2014}, the estimate of uncertainty allows for a choice of new training points to minimise the variance in some region of interest, with the potential of a large reduction in overhead. Algorithms such as those in \cite{Wilkinson14} and \cite{MeedsW14} which use this approach target an approximation to the true posterior and an accurate GP approximation is essential for the algorithm to be useful. By design, our algorithm overlays a standard MH or PMMH algorithm and it automatically targets the true posterior; an inaccurate approximation reduces the mixing efficiency but does not invalidate the algorithm. Again, since our algorithm overlays a standard MH algorithm the true likelihood (or an unbiased estimate) must be evaluated at every accepted point and it is not immediately obvious how estimates of uncertainty might be used to circuvent this requirement while maintaining the true posterior as the target.

Finally, estimating 
the hyper-parameters of a GP is computationally very costly, and this 
estimation should be repeated as the training data set grows. 
For this reason and because of the difficulty in identifying $d(d+1)/2$ kernel 
range parameters, GP methods often use a diagonal covariance structure
for the kernel, limiting the flexibility. Our approach of using the
covariance matrix of the sample from an initial training run to
provide a map to a new parameter space where Euclidean distance is
appropriate has a similar flavour to the pragmatic approach of using
an initial training sample to fit the GP hyper-parameters and then
keeping them fixed; alternatively, see \cite{FGPR:2006} for a partial solution.


\section*{Acknowledgements}
The authors thank Krysztof Latuszynski for a very helpful discussion with regard
to Theorem 3, and the Associate Editor 
and referee for useful suggestions that have improved the clarity 
of the paper.

\bibliographystyle{apalike}
\bibliography{kdadap}

\begin{thebibliography}{}

\bibitem[Andrieu et~al., 2010]{andrieu10}
Andrieu, C., Doucet, A., and Holenstein, R. (2010).
\newblock {P}article {M}arkov chain {M}onte {C}arlo methods (with discussion).
\newblock {\em J. R. Statist. Soc. B.}, 72(3):1--269.

\bibitem[Andrieu and Roberts, 2009]{AndrieuR09}
Andrieu, C. and Roberts, G.~O. (2009).
\newblock The pseudo-marginal approach for efficient {Monte Carlo}
  computations.
\newblock {\em The Annals of Statistics}, 37:697--725.

\bibitem[Andrieu and Thoms, 2008]{AT2008}
Andrieu, C. and Thoms, J. (2008).
\newblock A tutorial on adaptive {MCMC}.
\newblock {\em Stat. Comput.}, 18(4):343--373.

\bibitem[Banterle et~al., 2015]{BGLR:2015}
Banterle, M., Grazian, C., Lee, A., and Robert, C.~P. (2015).
\newblock Accelerating {M}etropolis-{H}astings algorithms by delayed
  acceptance.
\newblock \texttt{http://arxiv.org/abs/1406.2660}.

\bibitem[Bentley, 1975]{Bentley:1975}
Bentley, J.~L. (1975).
\newblock Multidimensional binary search trees used for associative searching.
\newblock {\em Commun. ACM}, 18(9):509--517.

\bibitem[Bliznyuk et~al., 2008]{BliznyukRSRWM08}
Bliznyuk, N., Ruppert, D., Shoemaker, C., Regis, R., Wild, S., and Mugunthan,
  P. (2008).
\newblock Bayesian calibration and uncertainty analysis for computationally
  expensive models using optimization and radial basis function approximation.
\newblock {\em Journal of Computational and Graphical Statistics}, 17:270--294.

\bibitem[Boys et~al., 2008]{boys08}
Boys, R.~J., Wilkinson, D.~J., and Kirkwood, T. B.~L. (2008).
\newblock Bayesian inference for a discretely observed stochastic-kinetic
  model.
\newblock {\em Stat. Comput.}, 18:125--135.

\bibitem[Christen and Fox, 2005]{ChristenF05}
Christen, J.~A. and Fox, C. (2005).
\newblock Markov chain {M}onte {C}arlo using an approximation.
\newblock {\em Journal of Computational and Graphical Statistics}, 14:795--810.

\bibitem[Conrad et~al., 2014]{CMPS:2014}
Conrad, P.~R., Marzouk, Y.~M., Pillai, N.~S., and Smith, A. (2014).
\newblock Accelerating asymptotically exact {MCMC} for computationally
  intensive models via local approximations.
\newblock \texttt{http://arxiv.org/abs/1402.1694}.

\bibitem[Craiu et~al., 2015]{CGLMRR2015}
Craiu, R.~V., Gray, L., Latuszy'nski, K., Madras, N., Roberts, G.~O., and
  Rosenthal, J.~S. (2015).
\newblock Stability of adversarial {Markov} chains, with an application to
  adaptive {MCMC} algorithms.
\newblock {\em Ann. Appl. Probab.}, 25(6):3592--3623.

\bibitem[Cui et~al., 2011]{CuiFO11}
Cui, T., Fox, C., and O'Sullivan, M.~J. (2011).
\newblock Bayesian calibration of a large-scale geothermal reservoir model by a
  new adaptive delayed acceptance {M}etropolis {H}astings algorithm.
\newblock {\em Water Resources Research}, 47:W10521.

\bibitem[Fearnhead et~al., 2014]{fearnhead12}
Fearnhead, P., Giagos, V., and Sherlock, C. (2014).
\newblock Inference for reaction networks using the {L}inear {N}oise
  {A}pproximation.
\newblock {\em Biometrics}, 70:457--466.

\bibitem[Fielding et~al., 2011]{FieldingNL11}
Fielding, M., Nott, D.~J., and Liong, S.-Y. (2011).
\newblock Efficient {MCMC} schemes for computationally expensive posterior
  distributions.
\newblock {\em Technometrics}, 53:16--28.

\bibitem[Finkel and Bentley, 1974]{FinkelBentley:1974}
Finkel, R.~A. and Bentley, J.~L. (1974).
\newblock Quad trees - a data structure for retrieval on composite keys.
\newblock {\em Acta Informatica}, 4:1--9.

\bibitem[Friedman et~al., 1977]{Friedman:1977}
Friedman, J.~H., Bentley, J.~L., and Finkel, R.~A. (1977).
\newblock An algorithm for finding best matches in logarithmic expected time.
\newblock {\em ACM Trans. Math. Softw.}, 3(3):209--226.

\bibitem[Gillespie, 1977]{Gillespie77}
Gillespie, D.~T. (1977).
\newblock Exact stochastic simulation of coupled chemical reactions.
\newblock {\em J. Phys. Chem.}, 81:2340--2361.

\bibitem[Golightly et~al., 2015]{GolightlyHS15}
Golightly, A., Henderson, D.~A., and Sherlock, C. (2015).
\newblock Delayed acceptance particle {MCMC} for exact inference in stochastic
  kinetic models.
\newblock {\em Statistics and Computing}, 25:1039--1055.

\bibitem[Golightly and Wilkinson, 2005]{Golightly05}
Golightly, A. and Wilkinson, D.~J. (2005).
\newblock Bayesian inference for stochastic kinetic models using a diffusion
  approximation.
\newblock {\em Biometrics}, 61(3):781--788.

\bibitem[Golightly and Wilkinson, 2011]{golightly11}
Golightly, A. and Wilkinson, D.~J. (2011).
\newblock Bayesian parameter inference for stochastic biochemical network
  models using particle {M}arkov chain {M}onte {C}arlo.
\newblock {\em Interface Focus}, 1(6):807--820.

\bibitem[Hastie et~al., 2009]{HastieTF09}
Hastie, T., Tibshirani, R., and Friedman, J. (2009).
\newblock {\em The Elements of Statistical Learning: Data Mining, Inference,
  and Prediction}.
\newblock Springer, New York, second edition.

\bibitem[Higdon et~al., 2011]{HigdonRMVF11}
Higdon, D., Reese, C.~S., Moulton, J.~D., Vrught, J.~A., and Fox, C. (2011).
\newblock Posterior exploration for computationally intensive forward models.
\newblock In Brooks, S., Gelman, A., Jones, G.~L., and Meng, X.-L., editors,
  {\em {H}andbook of {M}arkov {C}hain {M}onte {C}arlo}, chapter~16, pages
  401--418. Chapman \& Hall/CRC, Boca Raton, FL.

\bibitem[Joseph, 2012]{Joseph12}
Joseph, V.~R. (2012).
\newblock Bayesian computation using {D}esign of experiments-based
  {I}nterpolation technique.
\newblock {\em Technometrics}, 54:209--225.

\bibitem[Joseph, 2013]{Joseph13}
Joseph, V.~R. (2013).
\newblock A note on nonnegative {DoIt} approximation.
\newblock {\em Technometrics}, 55:103--107.

\bibitem[Kennedy and O'Hagan, 2001]{KennedyO01}
Kennedy, M.~C. and O'Hagan, A. (2001).
\newblock Bayesian calibration of computer models (with discussion).
\newblock {\em JRSSB}, 63:425--464.

\bibitem[Koblents and Miguez, 2015]{koblents15}
Koblents, E. and Miguez, J. (2015).
\newblock A population {M}onte {C}arlo scheme with transformed weights and its
  application to stochastic kinetic models.
\newblock {\em Statistics and Computing}, 25(2):407--425.

\bibitem[{Kostov} and {Whiteley}, 2016]{KostovWhitely2016}
{Kostov}, S. and {Whiteley}, N. (2016).
\newblock {An algorithm for approximating the second moment of the normalizing
  constant estimate from a particle filter}.
\newblock \texttt{http://arxiv.org/abs/1602.02279}.

\bibitem[Meeds and Welling, 2014]{MeedsW14}
Meeds, E. and Welling, M. (2014).
\newblock {GPS-ABC}: {Gaussian} process surrogate approximate {Bayesian}
  computation.
\newblock In {\em Thirtieth Conference on Uncertainty in Artificial
  Intelligence (UAI)}.

\bibitem[Overstall and Woods, 2013]{OverstallW13}
Overstall, A.~M. and Woods, D.~C. (2013).
\newblock A strategy for {B}ayesian inference for computationally expensive
  models with application to the estimation of stem cell properties.
\newblock {\em Biometrics}, 69:458--468.

\bibitem[Owen et~al., 2015]{owen15}
Owen, J., Wilkinson, D.~J., and Gillespie, C.~S. (2015).
\newblock Scalable inference for {M}arkov processes with intractable
  likelihoods.
\newblock {\em Statistics and Computing}, 25(1):145--156.

\bibitem[Payne and Mallick, 2014]{PayneM14}
Payne, R.~D. and Mallick, B.~K. (2014).
\newblock Bayesian big data classification: a review with complements.
\newblock \texttt{http://arxiv.org/abs/1411.5653}.

\bibitem[Picchini, 2014]{picchini14}
Picchini, U. (2014).
\newblock Inference for {SDE} models via {A}pproximate {B}ayesian
  {C}omputation.
\newblock {\em Journal of Computational and Graphical Statistics},
  23(4):1080--1100.

\bibitem[Quiroz, 2015]{Quiroz15}
Quiroz, M. (2015).
\newblock Speeding up {MCMC} by delayed acceptance and data subsampling.
\newblock \texttt{http://arxiv.org/abs/1507.06110}.

\bibitem[Rasmussen, 2003]{Rasmussen03}
Rasmussen, C.~E. (2003).
\newblock Gaussian processes to speed up hybrid {M}onte {C}arlo for expensive
  {B}ayesian integrals.
\newblock In Bernardo, J.~M., Bayarri, M.~J., Berger, J.~O., Dawid, A.~P.,
  Heckerman, D., Smith, A. F.~M., and West, M., editors, {\em Bayesian
  Statistics 7}, pages 651--659, Oxford. Oxford University Press.

\bibitem[Roberts and Rosenthal, 2001]{roberts2001}
Roberts, G.~O. and Rosenthal, J. (2001).
\newblock Optimal scaling for various {M}etropolis-{H}astings algo- rithms.
\newblock {\em Statistical Science}, 16:351--367.

\bibitem[Roberts and Rosenthal, 2004]{RobertsRosenthal:2004}
Roberts, G.~O. and Rosenthal, J.~S. (2004).
\newblock General state space {M}arkov chains and {MCMC} algorithms.
\newblock {\em Probab. Surv.}, 1:20--71.

\bibitem[Roberts and Rosenthal, 2007]{RobertsRosenthal:2007}
Roberts, G.~O. and Rosenthal, J.~S. (2007).
\newblock Coupling and ergodicity of adaptive {M}arkov chain {M}onte {C}arlo
  algorithms.
\newblock {\em J. Appl. Probab.}, 44(2):458--475.

\bibitem[Roberts and Rosenthal, 2009]{RR09}
Roberts, G.~O. and Rosenthal, J.~S. (2009).
\newblock Examples of adaptive {MCMC}.
\newblock {\em J. Comput. Graph. Statist.}, 18(2):349--367.

\bibitem[Roberts and Tweedie, 1996]{RobertsTweedie:1996}
Roberts, G.~O. and Tweedie, R.~L. (1996).
\newblock Geometric convergence and central limit theorems for multidimensional
  {H}astings and {M}etropolis algorithms.
\newblock {\em Biometrika}, 83(1):95--110.

\bibitem[Sacks et~al., 1989]{SWMW89}
Sacks, J., Welch, W.~J., Mitchell, T.~J., and Wynn, H.~P. (1989).
\newblock Design and analysis of computer experiments.
\newblock {\em Statistical Science}, 4:409--435.

\bibitem[Shen et~al., 2006]{FGPR:2006}
Shen, Y., Ng, A.~Y., and Seeger, M. (2006).
\newblock Fast {G}aussian process regression using kd-trees.
\newblock In {\em In Advances in Neural Information Processing Systems 18}. MIT
  Press.

\bibitem[Sherlock et~al., 2010]{SFR10}
Sherlock, C., Fearnhead, P., and Roberts, G.~O. (2010).
\newblock The random walk {M}etropolis: linking theory and practice through a
  case study.
\newblock {\em Statist. Sci.}, 25(2):172--190.

\bibitem[Sherlock et~al., 2015a]{STG15}
Sherlock, C., Thiery, A., and Golightly, A. (2015a).
\newblock Efficiency of delayed acceptance random walk {M}etropolis algorithms.
\newblock {\em In prepartion}.
\newblock \texttt{http://arxiv.org/abs/1506.08155}.

\bibitem[Sherlock et~al., 2015b]{sherlock15}
Sherlock, C., Thiery, A., Roberts, G.~O., and Rosenthal, J.~S. (2015b).
\newblock On the efficiency of pseudo-marginal random walk {M}etropolis
  algorithms.
\newblock {\em The Annals of Statistics}, 43(1):238--275.

\bibitem[Storer, 2002]{Storer:2002}
Storer, J. (2002).
\newblock {\em An introduction to data structures and algorithms}.
\newblock Springer-Verlag, New York.

\bibitem[van Kampen, 2001]{kampen2001}
van Kampen, N.~G. (2001).
\newblock {\em Stochastic Processes in Physics and Chemistry}.
\newblock North-Holland.

\bibitem[Vihola, 2012]{Vihola2012}
Vihola, M. (2012).
\newblock Robust adaptive {M}etropolis algorithm with coerced acceptance rate.
\newblock {\em Stat. Comput.}, 22(5):997--1008.

\bibitem[Wilkinson, 2012]{Wilkinson12}
Wilkinson, D.~J. (2012).
\newblock {\em Stochastic Modelling for Systems Biology}.
\newblock Chapman and Hall/CRC Press, London, 2nd edition.

\bibitem[Wilkinson, 2014]{Wilkinson14}
Wilkinson, R.~D. (2014).
\newblock Accelerating {ABC} methods using {Gaussian} processes.
\newblock In {\em JMLR Workshop and Conference Proceedings Volume 33:
  Proceedings of the Seventeenth International Conference on Artificial
  Intelligence and Statistics}.

\end{thebibliography}

\clearpage


\appendix

\numberwithin{equation}{section}
\renewcommand\thefigure{\thesection.\arabic{figure}}    
\setcounter{figure}{0}    
\renewcommand\thetable{\thesection.\arabic{table}}    
\setcounter{table}{0}

\begin{center}
\textbf{\LARGE Supplementary material for Adaptive, delayed-acceptance MCMC for targets with expensive likelihoods}
\end{center}

\section{Operations on the KD-tree}
In this appendix we describe the standard operations of creating a
balanced KD-tree from a dataset and of searching through the KD-tree
for the K-nearest neighbours.

\subsection{Creating a balanced KD-tree from an initial dataset}
\label{sect.app.create.bala}
Here we suppose the availability of $n_0$ pairs of a co-ordinate vector
$\theta$, and a vector of interest, $v$. The algorithm for creating
the tree proceeds recursively until the number of leaves in the node is
 less than $2b$.

The recursive function commences at the root node with $d_{split}=1$,
and a data set $\mathcal{D}$, which is all of the data, so that $n=\Abs{\mathcal{D}}=n_0$.

\begin{enumerate}
\item If $n<2b$ then finish this recursion; this node is a leaf node
  and contains data $\mathcal{D}$ and splitting component $d_{split}$.
\item Otherwise $n\ge 2b$, so find the median
  $\mu_{d_{split}}$ of the $\Abs{\mathcal{D}}$ scalar values
  for $\theta_{d_{split}}$. 
\item Place all points for which $\theta_{d_{split}}<\mu_{d_{split}}$
  into a set  $\mathcal{L}$.
\item Place all points for which $\theta_{d_{split}}>\mu_{d_{split}}$
  into a set  $\mathcal{R}$.
\item If any points have $\theta_{d_{split}}=\mu_{d_{split}}$ then
  use independent Bernoulli trials with probability $0.5$ to
  place each in $\mathcal{L}$ or $\mathcal{R}$.
\item Create a left and a right child for this node. 
\item Recursively call this function for the left child using
  $\mathcal{D}=\mathcal{L}$ and $d_{split}=d_{split}\oplus 1$.
\item Recursively call this function for the right child using
  $\mathcal{D}=\mathcal{R}$ and $d_{split}=d_{split}\oplus 1$.
\end{enumerate} 

\subsection{Finding the $k$ nearest neighbours of a point}
\label{sect.app.search.tree}
The algorithm below finds the $k$ nearest neighbours to a point $\theta^*$.
For simplicity it, assumes that
$k\le b$; it is straightforward (but messy) to alter it to allow any $k$.

\textbf{Stage One} 
\begin{enumerate}
\item
Descend from the root node to find the leaf node to which $\theta^*$ would be added if
we wished to add a new entry $(\theta^*,l_{\theta^*})$. 
\item Find and store an
ordered list of the $k$ points on this leaf that are nearest to $\theta^*$. Keep track
of $r_{max}$, the distance of the $k^{th}$ furthest point from $\theta^*$.
\end{enumerate}
Stage Two is only applied if the leaf note reached by Stage One
  has a parent; this is the case in all but the trivial tree.

\textbf{Stage Two}:
This part of the algorithm is initialised with a list of $k$ nearest
neighbours and a value $r_{max}$, both
obtained from Stage One. It uses an [\texttt{ascending/descending}]
flag and commences at 
the parent of the leaf node reached by Stage One with the flag
 set to \texttt{ascending}. To aid the linguistic flow we use the verbs
 \texttt{ascend} and \texttt{descend} as shorthand for calling Stage
 Two with the flag set to \texttt{ascending} and \texttt{descending}
 respectively. If the current node is a branch, let $\theta_{split}$
 and $d_{split}$ refer to the current node.
\begin{enumerate}
\item If \texttt{ascending}:
\begin{itemize}
\item[(i)] If $\Abs{\theta_{split}-\theta^*_{d_{split}}}<r_{max}$ then set 
$\bmu=\theta_{split}\bme_{d_{split}}$, where $\bme_i$ is the
  k-vector with all elements set to zero  except for the $i$th, which
  is set to one, then \texttt{descend} to the child (of this node) from
  which the algorithm has \textit{not} just ascended.
\item[(ii)] If $\Abs{\theta_{split}-\theta^*_{d_{split}}}\ge r_{max}$ and 
  if this node has a parent then \texttt{ascend}
  to this parent.
\end{itemize} 
\item If \texttt{descending} and the current node is a leaf then
calculate the distance from every point on the
leaf to $\bmtheta^*$ and sort the leaves according to this
distance. If any are within $r_{max}$ then update the list of the $k$
nearest neighbours and update $r_{max}$.
\item If \texttt{descending} and the current node is a branch then if
  $\theta^*_{d_{split}}\le\theta_{split}$ define the \textit{near
  child} to be the left child, otherwise define it to be the right
child. Similarly,  define the \textit{far child} to be whichever child is not the
near child.
\begin{itemize}
\item[(i)] \texttt{Descend} to the near child with $\bmu$ unchanged.
\item[(ii)] Update $u_{d_{split}}$ to $\theta_{split}$, keeping the other
  elements of $\bmu$ unchanged. If
  $\Norm{\bmu}<r_{max}$ then \texttt{descend} to the
  far child.
\end{itemize}
\end{enumerate}

\section{Proofs of Propositions 1 and 2}
\label{sec.proof.boundpkeepEneps}
\subsection*{Proof of Proposition 1}
Let $B^*_{\epsilon}$  and
$B^i_{\epsilon}$ be the $\epsilon$ balls around $\theta^*$ and $\theta_i$
respectively and denote the empty set by $\phi$. Firstly, since
$\theta_i~(i=1,\dots,n)$ are identically distributed,
\[
1-\Prob{\theta_1\notin B^*_{\epsilon},\dots,\theta_n\notin B^*_{\epsilon}}
=\Prob{\theta_1\in B^*_{\epsilon}~\mbox{or}\dots\mbox{or}~\theta_n\in B_\epsilon}
\le n\Prob{\theta_1\in B_{\epsilon}}=E_{n,\epsilon}.
\]
Hence $p_{keep}\ge 1-E_{n,\epsilon}$. 

Secondly, requiring that one point be outside of the $\epsilon$ ball of
another point is equivalent to requiring that the $\epsilon/2$ balls
of both points do not intersect. So, whatever the relationship between $\theta_1,\dots,\theta_{i-1}$,
\begin{align*}
\Prob{\theta_{i}\notin
  B^*_{\epsilon}|\theta_1\notin B^*_{\epsilon},\dots,\theta_{i-1}\notin
  B^*_{\epsilon}}
&=
\Prob{B^i_{\epsilon/2}\cap 
  B^*_{\epsilon/2}=\phi|B^1_{\epsilon/2}\cap B^*_{\epsilon/2}=\phi,\dots,B^{i-1}_{\epsilon/2}\cap B^*_{\epsilon/2}=\phi}\\
&\le
\Prob{B^i_{\epsilon/2}\cap 
  B^*_{\epsilon/2}=\phi}\\
&= \Prob{\theta_i\notin B^*_{\epsilon}}
\end{align*}
 since the conditioned event 
 only rules out portions of 
$\mathbb{R}^d\backslash B^*_{\epsilon/2}$ for $B^i_{\epsilon}$. Hence
\begin{align*}
p_{keep}&=\Prob{\theta_1\notin B^*_{\epsilon}}\prod_{i=2}^n\Prob{\theta_i\notin
  B^*_{\epsilon}|\theta_1\notin B^*_{\epsilon},\dots,\theta_{i-1}\notin
  B^*_{\epsilon}}\\
&\le \Prob{\theta_1\notin B^*_{\epsilon}}^n=(1-E_{n,\epsilon}/n)^n\le e^{-E_{n,\epsilon}}.
\end{align*}

\subsection*{Proof of Proposition 2}
 In this case the
distance, $D$ between any two points chosen independently satisfies $D^2/2\sim\chi^2_d$. 
The $\theta_i,~(i=1,\dots,d)$ are not mutually independent since
none can be within $\epsilon$ of any other, but this is irrelevant to
the calculation that follows. Let
$N_{n,\epsilon}:=\sum_{i=1}^n\ind{\theta_i\in B(\theta^*,\epsilon)}$, then
\[
E_{n,\epsilon}=\sum_{i=1}^n\Prob{\theta_i\in B(\theta_*,\epsilon)}
=nF_{\chi^2_d}(\epsilon^2/2),
\]
as required.

\section{Adaptive-KD-tree, delayed-acceptance pseudo-marginal algorithm}
\label{app.psm.algo}
\textbf{Algorithm 2}: \textit{adaptive-KD-tree, delayed-acceptance,
  pseudo-marginal Metropolis-Hastings.}

At the start of iteration $n$, let the current parameter value be
$\theta$ and let the cheap approximation to the posterior be
$\pihat_a(\theta)$ with the expensive
 unbiased estimate of the posterior denoted by $\pihat_s(\theta)$.

\begin{enumerate}
\item With probability $\beta$ go to Step \ref{step.PMMH} (PsMMH)
  else go to
  Step \ref{step.DA} (da-PsMMH). 
\item \label{step.PMMH}
\textbf{PsMMH} 
Propose $\theta^*$ from $q(\theta^*|\theta)$. 
Evaluate $\pihat_s(\theta^*)$ by effectively proposing $W^*$ from $\tilde{q}(W^*|\theta^*)$. Accept the proposal ($\theta\leftarrow \theta^*$) with
  probability given by (9); otherwise reject the
  proposal ($\theta\leftarrow \theta$). Go to Step \ref{step.adapB}.
\item \textbf{da-PsMMH} Propose $\theta^*$ from $q'(\theta^*|\theta)$.
\label{step.DA}
\begin{enumerate}
\item
\label{step.StageOne} \textbf{Stage 1}:
Evaluate $\pihat_a(\theta^*)$ using the current KD-tree; with
probability $\tilde{\alpha}_1(\theta,\theta^*)$ as defined in
(2) (but with $q'$ instead of $q$)  proceed to
Step \ref{step.adapSt2} (Stage 2); otherwise reject the proposal
($\theta\leftarrow\theta$), set $i_n=i_{n-1}$ and go to next iteration. 
\item \textbf{Stage 2}: 
\label{step.adapSt2}
Evaluate $\pihat_s(\theta^*)$ by, effectively, proposing $W^*$ from
$\tilde{q}(W^*|\theta^*)$. Accept the proposal ($\theta\leftarrow \theta^*$) 
with probability $\tilde{\alpha}_{2,s}(\theta,\theta^*)$ as defined in
(10); otherwise reject the proposal ($\theta\leftarrow
\theta$). Go to Step \ref{step.adapB}.
\end{enumerate}
\item \label{step.adapB} Set $i_n=i_{n-1}+1$; add
  $\left(\theta^*,\pihat_s(\theta^*)\right)$ to a list of recently
  evaluated parameter/posterior pairs; with probability $p_{i_n}$ add
  all pairs from this list to the KD-tree and then remove all pairs from the
  list; go to next iteration.
\end{enumerate}

\section{Theorems 1,  2 and 3: discussion, statement and proofs}
\label{sect.prove.thrms}

We discuss Assumptions 1 and 2, then place our algorithms in the
more general framework in terms of which our proof of ergodicity is phrased. Next, we state and discuss Theorem 2 and then
show, as Theorem 3, 
that it is not always necessary to force the adaptation probabilities,
$p_i$ to tend to zero. 
Finally we prove Theorems 1, 2 and 3. For simplicity, for Theorems 1
and 2, we
consider only the daPsMMH algorithm since the daMH is a special case
of daPsMMH with $W=W^*=1$.

\subsection{Discussion of Assumptions 1 and 2}
\cite{RobertsRosenthal:2004} show that any Metropolis-Hastings
kernel on a compact state-space satisfies Assumption~1 provided
$\pi(\theta)$ is continuous and $q(\theta^*|\theta)$ is continuous and
positive. 
In the applications of interest to us all of the parameters are
positive and it is common practice
\cite[e.g.][]{Golightly05,picchini14,koblents15,owen15} 
to place a vague but proper uniform prior on the logarithm of each
parameter; furthermore, $\pi$ is continuous.
We may therefore choose $q$ so that Assumption~1 is satisfied. 
If $\Theta$ is
not compact then Assumption~1 still 
holds, for example, if the proposal density $q(\theta^*|\theta)=q(\theta^*)$ is
an independence proposal which satisfies 
$q(\theta^*)>\delta \pi(\theta^*)$ since then
$q(\theta^*|\theta)\alpha(\theta,\theta^*)>q(\theta^*)\wedge \delta \pi(\theta^*)=\delta\pi(\theta^*)$.

Whether or not Assumption~2 is satisfied depends upon the exact
pseudo-marginal method used. Suppose, as in our examples, that an
unbiased estimate of the likelihood, $\Phat(y|\theta)$, (and hence, up to a constant, of
the posterior) is obtained using a bootstrap particle filter. This provides
$\Phat(y|\theta)$ as a finite product of Monte Carlo averages of
likelihood terms, where each of these terms arises from an assumed
distribution for the error in the observation of some stochastic
process $Z_t$. Suppose, for example, that 
$Y_t|z_t \sim N(z_t,\theta_1^2)$ and that the state space for $Z_t$ is
bounded, so that $0\le a_t\le (y_t-z_t)^2\le b_t<\infty$. Then 
$\log W\le (b_t-a_t)/(2\theta_1^2)$ and  Assumption~2 will therefore hold provided the support for
$\theta_1$ is bounded away from zero.

\subsection{More general set up}
\label{sect.general.setup}
We set up 
the algorithm in more general terms since, in addition to the specific
algorithms we have described, our ergodicity result,
Theorem 1, is applicable to a general class of
adaptive pseudo-marginal algorithms and, to our knowledge, is the
first such result.

Let $\Ptil$ be a non-adaptive MH kernel on
$\Theta$ with proposal $q(\theta^*|\theta)$, and let
$\{\Ptil_\gamma\}_{\gamma\in\mathcal{G}}$ be  a (usually infinite) set of
 kernels on $\Theta$, where
$P_{\gamma}$ has
proposal $q_{\gamma}(\theta^*|\theta)$. All kernels are assumed to have
the same stationary density, $\pi(\theta)$.
At iteration
$n$ the adaptive kernel that will be used with probability $1-\beta$ is
$\Ptil_{\gamma_n}$ and,
 since the Markov chain is adaptive, let $\gamma_n$ be a realisation of the
random variable $\Gamma_n$, which depends on the history of the
chain. At iteration $n$, our kernel is
\begin{equation}
\label{eqn.gen.kernel}
\Ptil^*_{\gamma_n}=\beta \Ptil + (1-\beta)\Ptil_{\gamma_n}
~~~\mbox{for some}~\beta \in (0,1).
\end{equation}
In our case, each
$\Ptil_{\gamma},(\gamma\in\mathcal{G})$ uses
the same initial proposal $q'(\theta^*|\theta)$, and this proposal will typically differ from
the proposal, $q(\theta^*|\theta)$ used in the fixed kernel. 
 However, our proof of ergodicity
 applies in the more general set up of \eqref{eqn.gen.kernel}.

\subsection{Theorem \ref{thrm.decrease.beta}: choice of $n$ and $\beta$ as functions of $\alpha_1$}
\label{sect.choose.n.beta}
Consider a collection of kernels, indexed by $\xi\in\Xi$, each a
mixture of an adaptive kernel, indexed by $\xi$, and a common fixed kernel:
\begin{equation}
\label{eqn.gen.kernel.xi}
\Ptil^{*\xi}_{\gamma_n}=\beta \Ptil + (1-\beta)\Ptil^\xi_{\gamma_n}
~~~\mbox{for some}~\beta \in (0,1)~\mbox{and}~\gamma_n \in G^{\xi}.
\end{equation}
Let $\Ptil$ use a proposal $q$, such as that in (6).
 For a given value of $\xi$, all $\Ptil^\xi_{\gamma},~\gamma\in G^{\xi}$
 use the same proposal, $q^\xi$; this could be the
RWM proposal defined in (6), for instance. 

\begin{theorem}
\label{thrm.decrease.beta}
Consider a set of adaptive daPsMRWM Algorithms indexed by $\xi$, as
described in and around \eqref{eqn.gen.kernel.xi}. Let the common
fixed kernel, $\Ptil$ satisfy Assumption~1 and
let all
of the adaptive kernels have the same stationary density as $\Ptil$,
given in $\mathrm{(8)}$; in particular, all use the
same mechanism for generating $W$ which satisfies Assumption~2. 
In addition to $\mathrm{(5)}$, let
$\{p_i\}_{i\in \mathbb{N}}$ be a strictly decreasing
sequence. For each $\xi$, after $n$ iterations
 let $I^\xi_n$ be the number of evaluations of the expensive
posterior and  let $\alphabar_{1,n}^\xi$ be the average Stage One
acceptance rate. Assume that given
$\epsilon>0~ \exists~i_{crude}$ such that for all $n$ with 
$\Expect{I^\xi_n}>i_{crude}$ and
all $\xi$, 
\[
\Prob{\alphabar_{1,n}^\xi<2\alphabar^\xi}>1-\epsilon.
\] 
For some fixed $\kappa>0$, we then set
\begin{equation}
\label{eqn.alpha.to.beta}
\beta^\xi=\kappa \alphabar^\xi
\end{equation}

Subject to the above $\Expect{I^\xi_{n}}\rightarrow \infty$ as
$n\rightarrow \infty$. 
Further, for any $\epsilon>0$ and all $\xi\in\Xi$
it is possible to choose a single $E>0$ so that if 
$\Expect{I^\xi_{n}}>E$, for some $n$, then
the TVD between the Markov chain and $\pi$ after $n$ iterations is bounded by $\epsilon$.
\end{theorem}

Thus, provided that 
after a certain number of
expensive iterations the acceptance rate is typically no more than a
constant multiple of the long term average, $\alphabar^\xi$, it is
safe to set the probability of choosing the fixed kernel proportional
to $\alphabar^\xi$.

As the chain progresses and the number of expensive evaluations of the posterior increases, 
the representation of the posterior by the KD-tree improves and
$\alphabar_{1,n}$ initially increases (with a typical relative change of around 5-10\%) before settling down. 
In practice, we find the condition on the convergence of the
Stage 1 acceptance probabilities to be a reasonable assumption. 

\subsection{Diminishing probabilities of adaptation}
\label{sect.dimin.adap.fine.state}
After every expensive iteration, Algorithms 1 and 2 store the newly-evaluated
expensive posterior; they ensure that the diminishing-adaptation
condition (see Theorem \ref{thrm.RobRosb}) is
satisfied by, after the $i$th expensive iteration, adding all stored
expensive posteriors to the KD-tree with probability 
$p_i\rightarrow 0$. Tables~2 and 3 suggest that the algorithm is more
efficient the more slowly $p_i\downarrow 0$. The intuition behind
this, and the possible consequences for higher-dimensional systems if
$p_i$ were to remain bounded away from $0$ are
discussed in Section~5; the concern is that such an algorithm
might not be ergodic. In Appendix \ref{prove.diminish} we prove
the following.

\begin{theorem}
\label{thrm.diminish.fine}
Let $\pi(\theta)$ be a continuous density with respect to Lebesgue
measure on a hyper-rectangular state space, $\Theta$, and consider an
adaptive, delayed-acceptance MH algorithm the same as Algorithm 1
but
with the following alterations: 
$\forall i\in\mathbb{N}$, $p_{2i}=1$ and $p_{2i+1}=0$, 
and we remove the restriction on the
growth of the tree described in Section~2.6. If the kernel
satisfies the minorisation condition, Assumption 1, then the
diminishing adaptation condition is automatically satisfied.
\end{theorem}

\subsection{Proofs of Theorems 1, 2, and 3}
\label{sect.theory.proofs}

Throughout this section the total variation distance (TVD) between any two
probability measures, $\nu(\cdot)$ and $\pi(\cdot)$,
is denoted $\Norm{\nu-\pi}$. We require the following definition and two results.
\begin{definition}
Consider a Markov kernel $P$ on a statespace $\mathcal{X}$. A subset
$C\subseteq \mathcal{X}$ is \textit{small} if there exists a positive
integer $n_0$, $\epsilon>0$, and a probability measure $\nu(\cdot)$ on
$\mathcal{X}$ such that the following \textit{minorisation} condition
is satisfied
\begin{equation}
\label{eqn.minorisation}
P^{n_0}(x,\cdot)\ge \epsilon \nu(\cdot).
\end{equation}
\end{definition}
\begin{theorem}
\label{thrm.RobRosa}
\cite[][]{RobertsRosenthal:2004} Consider a Markov chain with invariant probability distribution
$\pi(\cdot)$. Suppose that the entire statespace is small
(i.e. \eqref{eqn.minorisation} is satisfied with $C=\mathcal{X}$). Then the chain is
uniformly ergodic, and in fact $\Norm{P^n(x,\cdot)-\pi(\cdot)}\le (1-\delta)^{[n/n_0]}$
\end{theorem} 
\begin{theorem}
\label{thrm.RobRosb}
\cite[][]{RobertsRosenthal:2007}
Consider an adaptive MCMC algorithm on a statespace $\mathcal{X}$ with
adaptive kernels $P_{\gamma}$, $\gamma\in\mathcal{G}$, and with $\pi(\cdot)$ stationary for
each $P_\gamma;~\gamma\in \mathcal{G}$. Under the following conditions
the adaptive algorithm is ergodic. 
\begin{enumerate}
\item (Simultaneous uniform ergodicity) For all $\epsilon>0$, there is
  $N=N(\epsilon)\in \mathbb{N}$ such that
  $\Norm{P_\gamma^N(x,\cdot)-\pi(\cdot)}_{TV}\le \epsilon$ for all
  $x\in \mathcal{X}$ and $\gamma\in \mathcal{G}$.
\item (Diminishing adaptation) For any $X_0=x_0$, $\Gamma_0=\gamma_0$,
\[
\sup_{x\in
  \mathbb{X}}\Norm{P_{\Gamma_{n+1}}(x,\cdot)-P_{\Gamma_n}\left(x,\cdot\right)}_{TV}
\cip 0,
\]
where the convergence in probability is with respect to the
distribution of $\Gamma_{n}$ and $\Gamma_{n+1}$ given $x_0$ and $\gamma_0$.
\end{enumerate}
\end{theorem}

\subsubsection{Proof of Theorem 1}
\label{sect.ergodicity}
We now show that a class of adaptive Metropolis-Hastings algorithms,
which includes our algorithms, is ergodic subject to Assumptions~1 and 2. 
In line with our KD-tree approximation, we assume that the target, $\pi$, and the
proposals, $q$, $q'$ and $q_\theta~(\theta\in\Theta)$, are all densities
with respect to Lebesgue measure. 
 
Any delayed-acceptance algorithm is simply an accept/reject
Markov chain with a non-standard acceptance
probability. The key point is that detailed balance is preserved
since (for the pseudo-marginal version)
\begin{multline*}
\pi(\theta)q_{\theta}(w)w~q(\theta^*|\theta)q_{\theta^*}(w^*)
~\tilde{\alpha}_1(\theta,\theta^*)\tilde{\alpha}_{2,PM}([\theta,w],[\theta^*,w^*])\\
=
\pihat_c(\theta)q(\theta^*|\theta)\tilde{\alpha}_1(\theta,\theta^*)\times 
q_{\theta}(w)q_{\theta^*}(w^*)
\times \frac{\pi(\theta)w }{\pihat_c(\theta)}\tilde{\alpha}_{2,PM}([\theta,w],[\theta^*,w^*]),
\end{multline*}
and each of the three terms in the product is invariant to $(\theta,w)\leftrightarrow(\theta^*,w^*)$. 
We therefore prove ergodicity (subject to conditions) for any adaptive
pseudo-marginal algorithm of the form given in and above \eqref{eqn.gen.kernel}.

We require the following components. Here 
$A$ and $\Atil$ denote any (Lebesgue) measurable subsets of
$\Theta$ and $\Theta\times \mathcal{W}$ respectively, and $\delta$
represents the Dirac delta function.
\begin{enumerate}
\item
A fixed pseudo-marginal kernel on $\Theta\times \mathcal{W}$ with
stationary density $\pi(\theta)q_{\theta}(w)w$:
\begin{multline*}
\Ptil\left([\theta,w],\Atil\right)=\left(1-\alphabar_{PM}\left([\theta,w]\right)\right)\delta_A\left([\theta,w]\right)\\
+\int_{\Atil}\md \theta^*~\md w^*~q\left(\theta^*|\theta\right)q_{\theta^*}\left(w^*\right)\alpha_{PM}\left([\theta,w],[\theta^*,w^*]\right).
\end{multline*}
Here $\alphabar_{PM}([\theta,w])$ is the acceptance probability from the
current value: 
\[\alphabar_{PM}([\theta,w])=\int_{\Theta\times\mathcal{W}}
\md \theta^*~\md w^*~q(\theta^*|\theta)q_{\theta^*}(w^*)\alpha_{PM}([\theta,w],[\theta^*,w^*]).\]
\item
The corresponding fixed `ideal' kernel on $\Theta$ with
stationary density $\pi(\theta)$, 
\[
P(\theta,A)=(1-\alphabar_{MH}(\theta))\delta_A(\theta)
+\int_A\md \theta^*~q(\theta^*|\theta)\alpha_{MH}(\theta,\theta^*),
\]
from which we are unable to sample because $\pi(\theta)$ and
$\pi(\theta^*)$ are needed in order to evaluate $\alpha_{MH}$. Here
$\alphabar_{MH}(\theta)=\int_\Theta
\md\theta^*~q(\theta^*|\theta)\alpha_{MH}(\theta,\theta^*)$.
\item
A set of additional (pseudo-marginal) kernels on $\Theta\times \mathcal{W}$:
$\{\Ptil_{\gamma}([\theta,w],\cdot)\}_{\gamma \in
  \mathcal{G}}$, as described in and above \eqref{eqn.gen.kernel}, and
with the same stationary density as $\Ptil$.
\item
A sequence of probabilities $p_n$ satisfying (5).
\end{enumerate}

The generic algorithm is then:

\textbf{Algorithm 2b}: \textit{generic, adaptive, pseudo-marginal, propose and accept/reject algorithm.}

Iteration $n$ commences with current value $[\theta^{(n)},w^{(n)}]$ and kernel
index $\gamma_n$ and involves the following two steps.

\begin{enumerate}
\item
Sample $[\theta^{(n+1)},w^{(n+1)}]$ from $\Ptil^*_{\gamma_n}$ as
defined in \eqref{eqn.gen.kernel}.
\item
With probability $p_n$ update $\gamma_n$ (i.e. update the adaptive
kernel) by 
including all relevant information obtained since the kernel was last
updated to create a new kernel.
\end{enumerate}

\begin{theorem}
\label{thrm.adap.ergod.gen}
Subject to Assumptions 1 and 2, Algorithm 2b is
 ergodic.
\end{theorem}

\textbf{Proof} The condition (5) ensures 
that the diminishing adaptation condition of Theorem \ref{thrm.RobRosb} is satisfied.

We next show that subject to Assumption~1 , $\Ptil$ satisfies a
similar condition to Assumption~1 and hence so
does each of the kernels in \eqref{eqn.gen.kernel}.
 This is then shown to ensure
simultaneous uniform ergodicity.

We first define 
\[
\tilde{\nu}(\theta,w):=\nu(\theta)\frac{1}{c}q_{\theta}(w)w,
\]
which is a density by (7); we refer to the
corresponding measure as $\tilde{\nu}(\cdot)$.
From \eqref{eqn.PM.alpha.ineq},
\[
q(\theta^*|\theta)q_{\theta^*}(w^*)\alpha_{PM}\left([\theta,w],[\theta^*,w^*]\right)
>q(\theta^*|\theta)q_{\theta^*}(w^*)\frac{w^*}{\wbar}\alpha_{MH}(\theta,\theta^*)
=\frac{c\delta}{\wbar}\tilde{\nu}(\theta^*,w^*).
\]
This implies that for any $\gamma$ and any measurable set $\Atil\in \Theta \times [0,\wbar]$,
\[
P^*_{\gamma}([\theta,w],\cdot) \ge \frac{c\beta\delta}{\wbar}\tilde{\nu}(\cdot).
\]
Hence, the entire-statespace ($\Theta \times [0,\wbar]$)
 is small with $n_0=1$ and 
\begin{equation}
\label{eqn.define.deltatil}
\epsilon=\deltatil:=c\beta\delta/\wbar;
\end{equation}
$\deltatil<1$ since $c/\wbar\le 1$.
Each kernel therefore individually satisfies the condition of Theorem
\ref{thrm.RobRosa} and hence  $N=[\log \epsilon
/(1-\deltatil)]+1$ ensures that the collection of all kernels
$P^*_{\gamma}$ satisfy Condition 1 of Theorem \ref{thrm.RobRosb}. $\blacksquare$

Although the kernel in Algorithm 2b is more general than that in
Algorithm 2, in one particular sense Algorithm 2 is not a special
case of  Algorithm 2b, since the former potentially updates the kernel only after each
expensive evaluation rather than after each iteration. Adaptation
times enter the proof of
Theorem \ref{thrm.adap.ergod.gen} through the
\textit{diminishing adaptation} condition (see Theorem \ref{thrm.RobRosb} of
this article), for which it suffices that the probability of a change
in the kernel at any given iteration, $n$, tends to zero as
$n\rightarrow \infty$.
For Algorithm 2 this is guaranteed through 
Condition (5); however it also holds for Algorithm 1 since 
$\lim_{n\rightarrow\infty}i_n=\infty$ almost surely, as we now demonstrate. Subject to Assumption~2, the acceptance probability for the fixed kernel
between $[\theta,w]$ and $[\theta^*,w^*]$ is
\begin{equation}
\label{eqn.PM.alpha.ineq}
\alpha_{PM}\left([\theta,w],[\theta^*,w^*]\right)
\ge
\alpha_{MH}\left(\theta,\theta^*\right)\left(1\wedge \frac{w^*}{w}\right)
\ge
\frac{w^*}{\wbar}\alpha_{MH}(\theta,\theta^*).
\end{equation}
By (7), the average acceptance rate from $[\theta,w]$ for the
fixed kernel is
\[
\alphabar_{PM}\left([\theta,w]\right)\ge
\frac{c}{\wbar}\Expects{\theta^*}{\alpha_{MH}(\theta,\theta^*)} 
=\frac{c\alphabar_{MH}(\theta)}{\wbar}.
\] 
The idealised Metropolis-Hastings kernel $P(\theta,\cdot)$ is
uniformly ergodic by Assumption 1 (\cite{RobertsRosenthal:2004}) so $\alphabar_{MH}(\theta)$ is bounded below by
some $\alpha_0>0$ (\cite{RobertsTweedie:1996} Proposition 5.1). Hence the overall acceptance rate is bounded
below by $\beta\alpha_0c/\wbar$,
and by the strong law of large numbers,
$\lim_{n\rightarrow \infty}i_n=\infty$.

\subsubsection{Proof of Theorem \ref{thrm.decrease.beta}}
\label{sect.choosebeta.prove}
The proof of ergodicity of adaptive MCMC algorithms in Theorem 5 of
\cite{RobertsRosenthal:2007} relies on 
 a hypothetical Markov chain (for us, $x':=\{[\theta',w']_i\}_{i\in \mathbb{N}}$), which is identical to the real chain up
 until some iteration $n_0$ and then continues in parallel with the real
 chain using the kernel at $n_0$ without any further
 adaptation. The kernels for this chain are $\Ptil_{\gamma'_n}$, where
\[
\gamma'_n=\left\{
\begin{array}{ll}
\gamma_n&n\le n_0\\
\gamma_{n_0}&n> n_0.
\end{array}
\right.
\]
After iteration $n_0$, 
the hypothetical chain clearly has $\pi$ as its
stationary distribution. Theorem \ref{thrm.RobRosa} then informs us that
after a further $n_1$ iterations 
\begin{equation}
\label{eqn.our.TVD}
\Norm{\Ptil^{n_1}_{\gamma_{n_0}}(x,\cdot)-\pi(\cdot)}\le (1-\deltatil)^{n_1},
\end{equation}
for any $x$, including the $[\theta,w]$ value of the chain after $n_0$ 
iterations; the hypothetical chain is close to the target. Diminishing adaptation is then used to show
that if $n_0$ is large enough then after these further $n_1$
iterations the real chain is close to the  hypothetical chain.

We use the same approach as in \cite{RobertsRosenthal:2007} and show
that for a given required TVD between the true chain and $\pi$, the
required run length can be specified in terms of the expected number
of expensive iterations. Since Stage Two acceptance probabilities are
irrelevant to our argument $\alpha_n^\xi$ and $\alphabar_n^\xi$ henceforth denote, respectively, the
Stage One acceptance probability and its average after $n$ iterations.
Throughout this proof, for simplicity of notation, we will suppress the superscript $\xi$ from the
terms $\alpha_n^{\xi}$, $\alphabar_n^{\xi}$, $\alphabar^{\xi}$,
$n_0^{\xi}$, $n_1^{\xi},I^\xi,J^\xi$. The point is that the rate of the
convergence in \eqref{eqn.diminish.expect} and the bound in
\eqref{eqn.ensure.satisfy.one} only depend
on $\xi$ through the expected number of expensive evaluations,
$I^\xi$. 

Let $I_{n}$ and $J_{n}$ be, respectively, the
number of expensive evaluations of the true algorithm between iterations $1$ and $n$ and
between iterations $n_0+1$ and $n_0+n$. First
we define the following events:
\begin{equation}
\label{eqn.events}
\mathcal{A}_n:=\{\alphabar_n<2\alphabar\}
~~~\mbox{and}~~~
\mathcal{B}_{n_0,n_1}:=\{1-(1-p_{I_{n_0}})^{J_{n_1}}<\epsilon\}.
\end{equation}

The probability that an iteration involves an
evaluation of the expensive posterior is
\[
\rho_n := \beta+(1-\beta)\alpha_n = 
\alphabar\left(\kappa+(1-\kappa\alphabar)\frac{\alpha_n}{\alphabar}\right). 
\]
Thus
\begin{equation}
\label{eqn.expect.In}
\Expect{I_n}=\sum_{j=1}^n\rho_j=n\alphabar\left(\kappa+(1-\kappa\alphabar)\frac{\alphabar_n}{\alphabar}\right).
\end{equation}
So $\Expect{I_n}>n\alphabar\kappa$ and 
$\Expect{I_n}\rightarrow \infty$. 
Now, $\mbox{Var}[I_n-I_{n-1}] = \rho_n(1-\rho_n)\le\rho_n$ and 
whether or not each iteration is expensive is a sequence of
independent Bernoulli trials, so $\mbox{Var}[I_n]\le \Expect{I_n}$. 
Combined with Chebyshev's inequality shows that as 
$\Expect{I_n}\rightarrow \infty$, $I_n\rightarrow
\infty$ 
in 
probability and hence that the diminishing adaptation
probabilities satisfy
\begin{equation}
\label{eqn.diminish.expect}
\Expect{p_{I_n}}\rightarrow 0.
\end{equation}

By assumption, for $\Expect{J_n}>i_{crude}$,
$\Prob{\mathcal{A}_n}>1-\epsilon$. Further, from \eqref{eqn.expect.In}, conditional on
$\mathcal{A}_n$, $\Expect{J_n} < n\alphabar\left(\kappa+2\right)$. Thus
\begin{equation}
\label{eqn.n.lower}
\mbox{Conditional on }\mathcal{A}_n,~n\alphabar>\frac{\Expect{J_n}}{\kappa+2}.
\end{equation}
The TVD between the true adaptive chain and the hypothetical chain described in the preliminaries for
this proof is bounded above by the
probability that they are not coupled
\cite[e.g][]{RobertsRosenthal:2004}, and 
 since adaptation can only occur after the true
posterior (or an unbiased estimate thereof) is evaluated, the TVD
between the real chain and the hypothetical chain after $n_0+n_1$
iterations is less than $1-(1-p_{I_{n_0}})^{J_{n_1}}$.
Given \eqref{eqn.our.TVD}, the triangle inequality and the
monotonicity of $p_i$,
the following two conditions, therefore,
guarantee that the TVD between the true chain and
$\pi$ is less than $2\epsilon$: $\mathcal{B}_{n_0,n_1}$ and
\begin{align}
\label{eqn.cond.one.cvg.tev}
(1-\deltatil)^{n_1}&< \epsilon.
\end{align}
By \eqref{eqn.alpha.to.beta} and \eqref{eqn.define.deltatil},
$\deltatil=\alphabar c\kappa \delta/\wbar<1$ since
$c/\wbar\le 1$. Also $\log(1-\deltatil)< -\deltatil$. Thus,
\[
n_1\log(1-\deltatil)<-n_1\deltatil=-\frac{n_1\alphabar c\kappa\delta}{\wbar}
<-\Expect{J_{n_1}}\frac{c\kappa\delta}{\wbar(\kappa+2)},
\]
by \eqref{eqn.n.lower} applied to the iterations from $n_0+1$.
Hence, conditional on $\mathcal{A}_{n_1}$, \eqref{eqn.cond.one.cvg.tev} can be guaranteed by fixing
$\mathbb{E}[J_{n_1}]$ (i.e. choosing $n_1$) 
\begin{equation}
\label{eqn.ensure.satisfy.one}
\Expect{J_{n_1}}>-\frac{\wbar(\kappa+2) \log \epsilon}{c\kappa\delta}.
\end{equation}
To deal with $\mathcal{B}_{n_0,n_1}$, for this fixed
$\Expect{J_{n_1}}$, apply Jensen's inequality twice and then \eqref{eqn.diminish.expect}:
\[
\Expect{(1-p_{I_{n_0}})^{J_{n_1}}}
\ge
\Expect{(1-p_{I_{n_0}})^{\Expect{J_{n_1}}}}
\ge
\Expect{1-p_{I_{n_0}}}^{\Expect{J_{n_1}}}
\rightarrow 1
\]
as $\Expect{I_{n_0}}\rightarrow \infty$. Hence, by Markov's inequality, for sufficiently large $\Expect{I_{n_0}}$ we can ensure $\Prob{\mathcal{B}_{n_0,n_1}}>1-\epsilon$.

With these choices of $\Expect{I_{n_0}}$ and $\Expect{J_{n_1}}$, and
hence of $E=\Expect{I_{n_0}}+\Expect{J_{n_1}}$, $\mathcal{A}_{n_1}$ and
$\mathcal{B}_{n_0,n_1}$ each holds with probability
$1-\epsilon$,  so the 
TVD between the true chain and $\pi$ is less than $4\epsilon$.

\subsubsection{Proof of Theorem \ref{thrm.diminish.fine}}
\label{prove.diminish}
We denote the k-nearest neighbour approximation to the posterior at $\theta^*$
after $n$ iterations by $\pihat_n(\theta^*)$ and a ball of radius $r$
centred at $\theta$ by $B_r(\theta)$. We denote 
the Stage One and Stage Two
acceptance probabilities at iteration $n$ by $\alpha_{1,n}(\theta,\theta^*)$ and
$\alpha_{2,n}(\theta,\theta^*)$.

For $\theta\in\Theta$ let 
\[A_n(\theta,r) :=\{\mbox{after $n$ iterations all $k$
nearest neighbours to $\theta$ lie within $B_{r}(\theta)$}\},\]
and note that $A_{n-1}(\theta,r)\Rightarrow A_n(\theta,r)$. 
If $A_n(\theta,r)$ holds then then
any change (adaptation) in the cheap estimate $\pihat_n(\theta)$ must occur
through one or more new points being added to the tree inside
$B_r(\theta)$.
Since $\log \pi$ is continuous, for any $\epsilon>0~\exists$ 
$\epsilon_\circ$ such that if  $A_{n-1}(\theta,\epsilon_{\circ})$ and  $A_{n-1}(\theta^*,\epsilon_{\circ})$
hold, then 
\[
1-\epsilon<\frac{\pihat_{n}(\theta)}{\pihat_{n-1}(\theta)}<1+\epsilon
~~~\mbox{and}~~~
1-\epsilon<\frac{\pihat_{n+1}(\theta^*)}{\pihat_{n}(\theta^*)}<1+\epsilon.
\]
$P_{\gamma_n}$ and $P_{\gamma_{n+1}}$ differ only in
 their acceptance probabilities,
with $P_{\gamma_n}$ using the
ratio of
$\pihat_{n}(\theta^*)$ and $\pihat_{n-1}(\theta)$ in both $\alpha_{1,n}$
and $\alpha_{2,n}$. Yet, subject to 
$A_{n-1}(\theta,\epsilon_{\circ})$ and  $A_{n-1}(\theta^*,\epsilon_{\circ})$,
\[
(1-\epsilon)^4 <
\frac{\alpha_{1,n+1}(\theta,\theta^*)\alpha_{2,n+1}(\theta,\theta^*)}
{\alpha_{1,n}(\theta,\theta^*)\alpha_{2,n}(\theta,\theta^*)}
<(1+\epsilon)^4.
\]
Since $\Theta$ is compact, $\log \pi$ is
uniformly continuous. 
Hence, for small enough $\epsilon$, and subject to 
$\cap_{\theta\in\Theta}A_{n-1}(\theta,\epsilon_{\circ})$,
\[||P_{\gamma_{n+1}}(\theta,\cdot)-P_{\gamma_n}(\theta,\cdot)||<5\epsilon.\]
  We
 show, given any $\epsilon>0$, $\exists~n_{\epsilon}$ such that
 $\Prob{\cap_{\theta^*\in\Theta}A_n(\theta^*,\epsilon_{\circ})}>1-2\epsilon$,
 for all $n\ge n_{\epsilon}$.  
 So for $n>n_{\epsilon}$, 
$\sup_{\theta}||P_{\gamma_{n+1}}(\theta,\cdot)-P_{\gamma_n}(\theta,\cdot)||<7\epsilon$.

First, partition the (hyperrectangular) state space, $\Theta$, into
$n_{\square}$ hypercubes of size $\epsilon_{\square}$, such that any
ball of radius $\epsilon_{\circ}$ must contain at least one
hypercube. We now place $n$ points uniformly at random in
$\Theta$ (i.e. according to a homogeneous Poisson process, $U$). Denote the
number of points that fall in the $i$th hypercube by $\square_i$. For
any
fixed $k\in\mathbb{N}$, 
$\Prob{\exists~i \in\{1,\dots,n_{\square}\}\mbox{ such that  }\square_i<k}\le n_{\square}\Prob{\square_1<k}\rightarrow 0$
as $n\rightarrow \infty$. Hence there is an $n_{\bullet}$ such that
for all $n\ge n_{\bullet}$, 
$\Prob{\exists~i \mbox{ such that  }\square_i<k|n~\mbox{points from}~U}<\epsilon$.

The minorisation condition holds over the whole
state space, and each kernel $P_{\gamma}$ is reversible; moreover
adaptation only occurs on even-numbered expensive iterations so that each kernel
is used (at least) twice before adaptation. 
Lemma 26 of \cite{CGLMRR2015} then implies that for each
pair of iterations there is a probability of at least
$\beta^2\delta^2/4$ of sampling from $\pi$.
Since $\log \pi$ is continuous and $\Theta$ is compact, 
\[
\rho:=\frac{\min_{\theta \in\Theta}\pi(\theta)}{\max_{\theta \in\Theta}\pi(\theta)}>0
\]
So whenever a sample from $\pi$ is obtained, a sample from the
homogeneous Poisson process, $U$, may be 
obtained with a probability of at least $\rho$. 
Define the event
\[
C_n:=\{\exists~\mbox{at least one sample from $U$ in $n$ iterations}\}.
\]
Then $\Prob{C_n^c}\le (1-\rho\beta^2\delta^2/4)^{n/2}$ and,
 given $\epsilon>0$ and $n_{\bullet}$
there is an $n_{once}$ such that for all $n\ge n_{once}$, 
$\Prob{C_n^c}<\epsilon/(n_\bullet)$. Hence for all $n\ge
n_{\epsilon}:=n_{once}n_{\bullet}$,
\[
\Prob{\mbox{after $n$ iterations}~\exists~i\in\{1,\dots,n_{\square}\}~\mbox{such that}~\square_i<k}
<2 \epsilon.
\]
If there are at least $k$ entries in each hypercube then, since all balls of
radius $r$ contain at least one hypercube, the $k$ nearest neighbours
to all $\theta^*\in\Theta$ must be within $B_r(\theta^*)$.

\section{Simulation study: model details and further results}
\subsection{Model and inference details and data simulation}
\label{app.model}
Tables~\ref{tab:tab1} and \ref{tab:tab2} list the 
reactions and associated hazards for the Lotka Volterra and
autoregulatory examples, respectively.

\begin{table}[t]
\centering
\begin{tabular}{@{}llll@{}}
  \hline
  Label  & Reaction &  Hazard & Description\\
  \hline
  $R_{1}$ & $\mathcal{X}_{1}\xrightarrow{\phantom{a}\nu_{1}\phantom{a}}  2\mathcal{X}_{1}$ & $\nu_{1}X_{1}$ & Prey reproduction\\
  $R_{2}$ & $\mathcal{X}_{1}+\mathcal{X}_{2}\xrightarrow{\phantom{a}\nu_{2}\phantom{a}}  2\mathcal{X}_{2}$ & $\nu_{2}X_{1}X_{2}$ & Prey death, predator reproduction\\
  $R_{3}$ & $\mathcal{X}_{2}\xrightarrow{\phantom{a}\nu_{3}\phantom{a}}  \emptyset$ & $\nu_{3}X_{2}$ & Predator death\\
  \hline
\end{tabular}\caption{Reaction list and hazards for the Lotka-Volterra system.}\label{tab:tab1}
\end{table} 
\begin{table}[t]
\centering
\begin{tabular}{@{}llll@{}}
  \hline
  Label  & Reaction &  Hazard & Description\\
  \hline
  $R_{1}$ & $\textsf{DNA}+\textsf{P}_2 \xrightarrow{\phantom{a}\nu_{1}\phantom{a}} \textsf{DNA}\cdot\textsf{P}_2$  & $\nu_{1}X_{1}X_{4}$ & Dimer binding\\
  $R_{2}$ & $\textsf{DNA}\cdot\textsf{P}_2 \xrightarrow{\phantom{a}\nu_{2}\phantom{a}} \textsf{DNA}+\textsf{P}_2$ & $\nu_{2}(k-X_{1})$ & Dimer unbinding\\
  $R_{3}$ & $\textsf{DNA} \xrightarrow{\phantom{a}\nu_{3}\phantom{a}} \textsf{DNA} + \textsf{RNA}$ & $\nu_{3}X_{1}$ & Transcription\\
  $R_{4}$ & $\textsf{RNA} \xrightarrow{\phantom{a}\nu_{4}\phantom{a}} \textsf{RNA} + \textsf{P}$ & $\nu_{4}X_{2}$ & Translation\\
  $R_{5}$ & $2\textsf{P} \xrightarrow{\phantom{a}\nu_{5}\phantom{a}} \textsf{P}_2$ & $\nu_{5}X_{3}(X_{3}-1)/2$  & Forward dimerisation\\ 
  $R_{6}$ & $\textsf{P}_2 \xrightarrow{\phantom{a}\nu_{6}\phantom{a}} 2\textsf{P}$ & $\nu_{6}X_{4}$ & Reverse dimerisation\\
  $R_{7}$ & $\textsf{RNA} \xrightarrow{\phantom{a}\nu_{7}\phantom{a}} \emptyset$ & $\nu_{7}X_{2}$  & RNA degradation\\
  $R_{8}$ & $\textsf{P} \xrightarrow{\phantom{a}\nu_{8}\phantom{a}} \emptyset$ & $\nu_{8}X_{3}$ & Protein degradation\\
  \hline
\end{tabular}\caption{Reaction list and hazards for the auto-regulatory system.}\label{tab:tab2}
\end{table} 

A single data set was simulated from the Lotka-Volterra MJP using an
initial value of $X_0=(71,79)$ and parameter values taken from \cite{Wilkinson12}, that is 
$\nu=(1.0,0.005,0.6)$. Each $X_t,~(t=1,\dots,50)$ was corrupted as in
equation (11) with
 $\sigma_{1}=\sigma_{2}=8$.

For the autoregulatory system we generated 2 synthetic datasets (labelled as $\mathcal{D}_{1}$ 
and $\mathcal{D}_{2}$) by taking $X_{0}=(5,8,8,8)$, $K=10$ and using 
parameter values taken from \cite{golightly11}, that is 
$\nu=(0.1,0.7,0.35,0.2,\linebreak[0]0.1,0.9,0.3,0.1)$. Simulated
values were corrupted with noise as in (11) with
 $\sigma_{1}=\sigma_{2}=0.5$ and $\sigma_{3}=\sigma_{4}=1$. Dataset 
$\mathcal{D}_{1}$ consists of 101 observations on the time interval $[0,100]$ 
and $\mathcal{D}_{2}$ consists of 201 observations on the time interval $[0,1000]$.

For the autoregulatory system, 
the total 
number $K$ of $\textsf{DNA}\cdot\textsf{P}_2$ and $\textsf{DNA}$ 
is fixed throughout the evolution of the system, for in our inferences
it is assumed to be known, 
so that the model comprises of 4 species. We denote 
the number of molecules of $\textsf{DNA}$, $\textsf{RNA}$, 
$\textsf{P}$ and $\textsf{P}_2$ as $X_{1}$, $X_{2}$, $X_{3}$ and 
$X_{4}$ respectively.
 As noted by 
\cite{golightly11}, the rate constants in the reversible reactions can be difficult to 
infer and we therefore fix $\nu_{1}$ and $\nu_{5}$ at the ground truth. We consider inference for 
the remaining rate constants and the observation error standard
deviations, $10$ parameters in all.  

\subsection{Tuning parameters for the fixed kernels}
\label{app.tuning.param}
For inference on the Lotka Volterra system, 
we follow the practical advice of \cite{sherlock15} by choosing the
number of particles, $m$, so that the variance ($\tau^{2}$) in 
the log-posterior at the median (estimated from the pilot run) and 4 additional sampled parameter values is less 
than $3$. We set $m=200$ since this gave $\tau^2\in[2.06,3.07]$. Under certain 
assumptions regarding target and the variance in the log-posterior, in
the case that the target is approximately Gaussian, \cite{sherlock15} found 
that the scaling of the proposal variance should be approximately $\lambda=(2.56^2/d)$ 
to optimise efficiency. We found further scaling this quantity by $1.1$ appeared to give 
optimal performance in terms of effective sample size (ESS) per second
giving $V_{fixed}=1.1\times(2.56^2/5)\times \hat{\Sigma}$.

For the autoregulatory example, following \cite{roberts2001}, we found
that scaling variance 
estimated from the tuning run by
$\lambda=(2.38^2/d)$ with $d=10$
gave optimal performance of the fixed kernel (ie without delayed acceptance).

The choices of $\epsilon=0.3065$ for the Lotka Volterra and
$\epsilon=0.982$ for the autoregulatory system followed from
considerations described immediately after (12).

\section{The linear noise approximation}
\label{app.lna}
The linear noise approximation (LNA) of the Markov jump process 
defined by the reaction hazards \cite[e.g.][]{fearnhead12} 
ignores discreteness (but not stochasticity) and gives the state 
$X_{t}$ as a Gaussian: $X_{t}\sim N\left(z_t+m_t\,,\,V_t\right)$,
where $z_t$, $m_t$ and $V_t$ satisfy a coupled ODE system
\begin{equation}\label{LNAode}
\left\{\begin{array}{rl}
\dot{z}_t =& \hspace{-0.2cm} S\,h(z_t,\nu) \\
\dot{m}_t =& \hspace{-0.2cm} F_t m_t\\
\dot{V}_t =& \hspace{-0.2cm} V_t F_{t}^{T}+S\textrm{diag}\left\{h(z_t,\nu)\right\}S^T +F_t V_t
\end{array}\right.
\end{equation}
Here, $h(x_t,\nu)$ is the length-8 column vector containing the reaction hazards, 
$F_t$ is a $4\times 4$ matrix whose $(i,j)$th entry is given by the first partial 
derivative of the $i$th component of $S\,h(z_t,\nu)$ with respect to the $j$th 
component of $z_t$ and $S$ is the $4\times 8$ stoichiometry matrix whose $(i,j)$th element gives the 
effect of reaction $j$ on species $i$.

\subsection{Marginal likelihood under the linear noise approximation}
\label{lna}

For simplicity of exposition we assume an observation regime of the form
\[
Y_{t}=X_{t}+\epsilon_{t}\,,\qquad \epsilon_{t}\sim \textrm{N}\left(0,\Sigma\right)
\] 
where $\epsilon_{t}$ is a length-$d_x$ Gaussian random vector and $t=0,1,\ldots,n$. Suppose that 
$X_{1}$ is fixed at some value $x_{1}$. The marginal likelihood $\pi(y_{1:n}|\theta)$ (and hence the 
posterior up to proportionality) under the LNA can be obtained as follows.
\begin{enumerate}
\item Initialisation. Compute 
\[
\pi(y_{1}|\theta)=\phi\left(y_{1}\,;\, x_{1}\,,\,\Sigma\right)
\]
where $\phi\left(y_{1}\,;\, x_{1}\,,\,\Sigma\right)$ denotes the Gaussian density 
with mean vector $x_{1}$ and variance matrix $\Sigma$. 
Set $a_{1}=x_{1}$ and $C$ to be the $d_{x}\times d_{x}$ matrix of zeros.
 
\item For times $t=1,2,\ldots ,n-1$,
\begin{itemize}
\item[(a)] Prior at $t+1$. Initialise the LNA with $z_{t}=a_{t}$, $m_{t}=0$ and $V_{t}=C_{t}$. 
Note that $m_{s}=0$ for all $s>t$. Integrate the ODE system (\ref{LNAode}) 
forward to $t+1$ to obtain $z_{t+1}$ and $V_{t+1}$. Hence
\[
X_{t+1}|y_{1:t},\theta\sim N(z_{t+1},V_{t+1})\,.
\]
\item[(b)] One step forecast. Using the observation equation, we have that 
\[
Y_{t+1}|y_{1:t},\theta\sim N\left(z_{t+1},V_{t+1}+\Sigma\right)\,.
\]
Compute
\begin{align*}
\pi(y_{1:t+1}|\theta)&=\pi(y_{1:t}|\theta)\,\phi\left(y_{t+1}\,;\, z_{t+1}\,,\,V_{t+1}+\Sigma\right)\,.
\end{align*}
\item[(c)] Posterior at $t+1$. Combining the distributions in (a) and (b) gives 
$X_{t+1}|y_{1:t+1},\theta\sim N(a_{t+1},C_{t+1})$ where
\begin{align*}
a_{t+1} &= z_{t+1}+V_{t+1}\left(V_{t+1}+\Sigma\right)^{-1}\left(y_{t+1}-z_{t+1}\right) \\
C_{t+1} &= V_{t+1}-V_{t+1}\left(V_{t+1}+\Sigma\right)^{-1}V_{t+1}\,.
\end{align*}

\end{itemize}
\end{enumerate}

\section{Additional graphics and discussion from the simulation study}
\label{sect.marginals}
\begin{figure}[ht]
\begin{center}
\psfrag{psi1}[][]{$\theta_{1}$}
\psfrag{psi2}[][]{$\theta_{2}$}
\psfrag{psi3}[][]{$\theta_{3}$}
\psfrag{psi4}[][]{$\theta_{4}$}
\psfrag{psi5}[][]{$\theta_{5}$}
\includegraphics[angle=270,width=\textwidth]{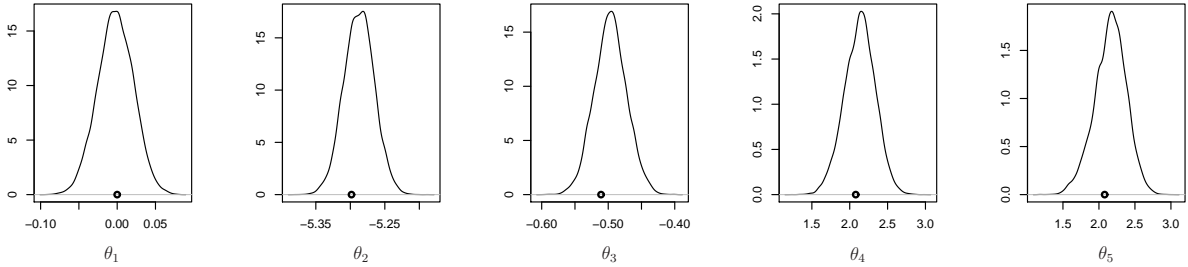}
\caption{Marginal posterior densities of $\theta_{i}$ ($i=1,\ldots,5$) based on the (thinned) output of da-PsMMH. \label{fig.LVdens}
}
\end{center}
\end{figure}

\begin{figure}[ht]
\begin{center}
\psfrag{AlogP}[][]{$\log(\hat{\pi}_{c}(\theta))$}
\psfrag{logP}[][]{$\log(\hat{\pi}_{s}(\theta))$}
\includegraphics[angle=270,width=\textwidth]{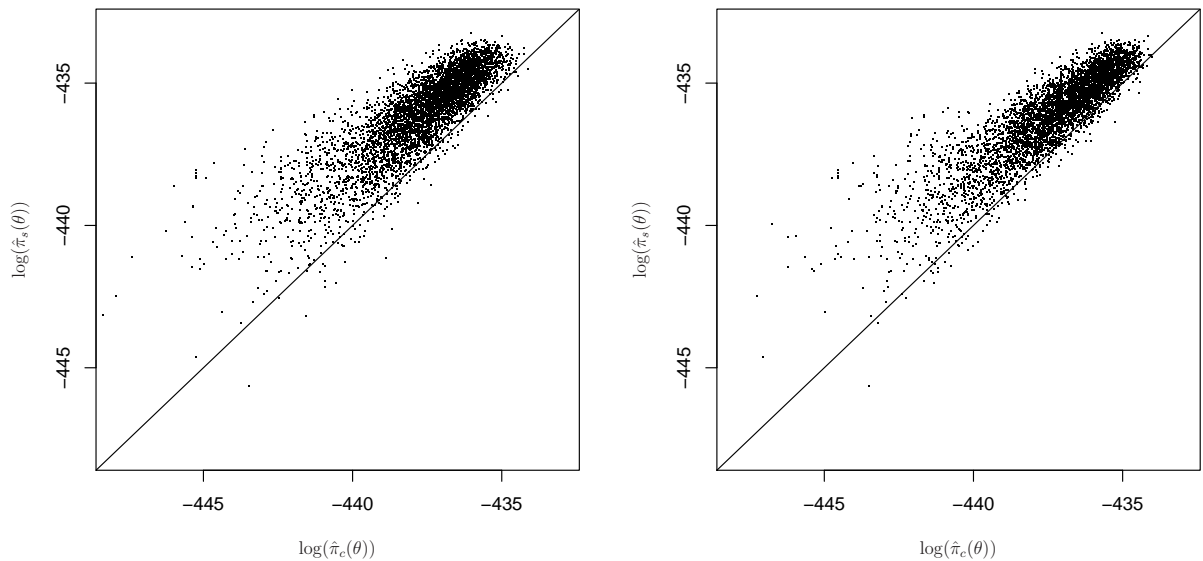}
\caption{Log-posterior estimates under the MJP ($\log(\hat{\pi}_{s}(\theta))$) against 
the corresponding log-posterior estimate given by the KD-tree ($\log(\hat{\pi}_{c}(\theta))$) based on 
the training data only (left panel) and the final adapted tree obtained after running da-PsMMH for $10^{5}$ seconds 
with $\xi=3$ and $k=b=10$ (right panel). Both plots are obtained using $5,000$ values of $\theta$ sampled from the posterior $\pi(\theta)$. \label{fig.LVem}}
\end{center}
\end{figure}

\begin{table}[ht]
\begin{center}
\begin{tabular}{|c|cccc|}
\hline
 & \multicolumn{4}{c|}{$2b$} \\
$k$ & 4  & 10 & 20 & 30 \\
\hline
2 &2807 (5.3) &2715 (5.1) &3435 (6.5) &3423 (6.5) \\
5 & &3017 (5.7) &3871 (7.3) &3848 (7.3) \\
10 & & &3591 (6.8) &3377 (6.4) \\
15 & & & &3330 (6.3) \\
\hline
\end{tabular}
\caption{Minimum effective sample size (mESS) and relative mESS in
  parentheses for the Lotka Volterra model.
\label{table.LVk}}
\end{center}
\end{table}
With regard to Table \ref{table.LVk}, 
we impose the restriction that 
$k\leq b$, since choosing $k>b$ automatically implies that the $k$ nearest neighbours 
to a particular parameter value will, at some point, be split over
more than one branch node. We found that using $k=2$ reduces the computational cost of searching the tree 
but also reduces the accuracy of the KD-tree approximation, resulting in an overall decrease 
in mESS. Similarly, for $k>5$ the increased accuracy is offset by increased computational 
cost. Using $k=5$ and $2b=20$ gave a 7-fold improvement in overall efficiency over PsMMH.

\begin{figure}[ht]
\begin{center}
\psfrag{psi1}[][]{$\theta_{1}$}
\psfrag{psi2}[][]{$\theta_{2}$}
\psfrag{psi3}[][]{$\theta_{3}$}
\psfrag{psi4}[][]{$\theta_{4}$}
\psfrag{psi5}[][]{$\theta_{5}$}
\psfrag{psi6}[][]{$\theta_{6}$}
\psfrag{psi7}[][]{$\theta_{7}$}
\psfrag{psi8}[][]{$\theta_{8}$}
\psfrag{psi9}[][]{$\theta_{9}$}
\psfrag{psi10}[][]{$\theta_{10}$}
  \includegraphics[angle=270,width=\textwidth]{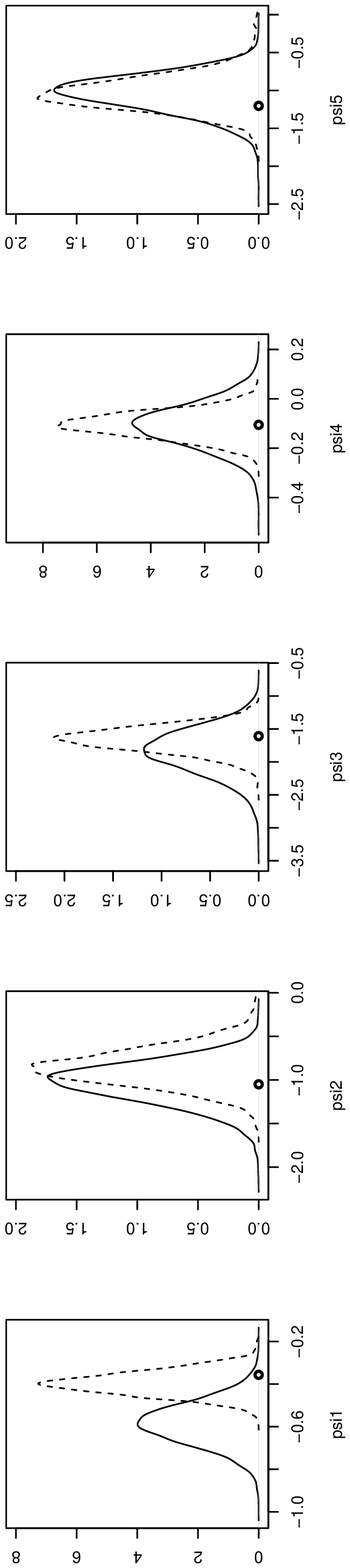}
 \includegraphics[angle=270,width=\textwidth]{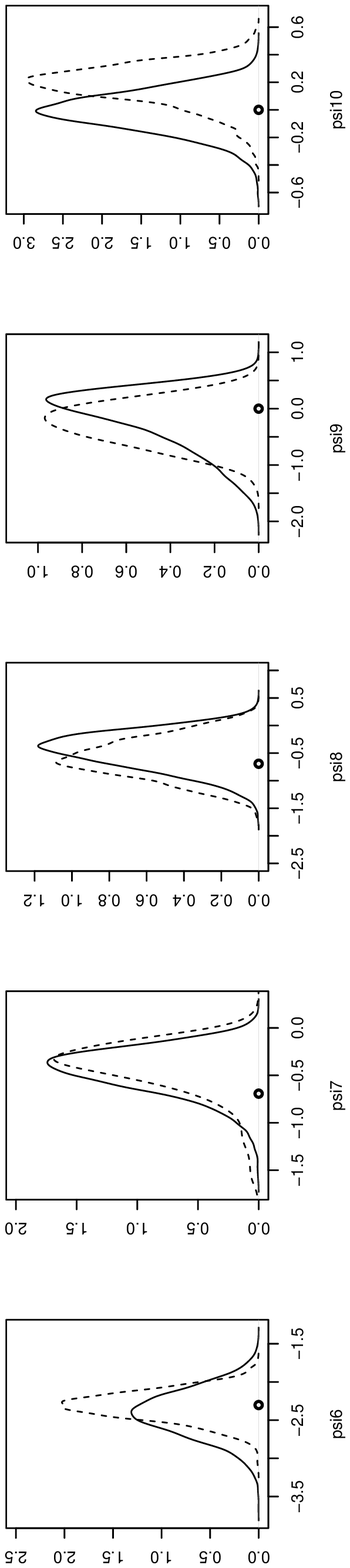}
\caption{Marginal posterior densities of $\theta_{i}$ ($i=1,\ldots,10$) based on the (thinned) output of da-MMH using dataset 
$\mathcal{D}_{1}$ (solid) and $\mathcal{D}_{2}$ (dashed). \label{fig.ARdens}
}
\end{center}
\end{figure}

\end{document}